\documentclass[lettersize,journal]{IEEEtran}                 %double column
 \usepackage{graphicx}
\usepackage{comment}
\usepackage{amsmath}
\usepackage{amsfonts}
\usepackage{bbding}
\usepackage{amssymb}
\usepackage{array}
\usepackage{multirow}
\usepackage{graphicx}
\usepackage[named]{algo}
\usepackage{algorithmic}
\usepackage{algorithm}
\usepackage{amsthm}
\usepackage{psfrag}
\usepackage[compress]{cite}

\usepackage{subfigure}
\usepackage{float}

\makeatletter
\renewcommand{\citepunct}{,\penalty\@m\hskip.13emplus.1emminus.1em}
\renewcommand{\citedash}{\hbox{--}\penalty\@m}
\makeatother
\usepackage{setspace}
\usepackage{color}
\allowdisplaybreaks

\usepackage{hyperref}
\usepackage{bm}
\usepackage{bbm}
\renewcommand{\IEEEQED}{\IEEEQEDopen}

\IEEEoverridecommandlockouts

\begin{document}

\title{Floor-Plan-aided Indoor Localization: Zero-Shot Learning Framework, Data Sets, and Prototype}

% \author{

% Haiyao Yu,
% Changyang She,
% Yunkai Hu,
% Rui Wang,
% Geng Wang,
% Branka Vucetic,~\IEEEmembership{Life Fellow,~IEEE}\\
% and Yonghui Li,~\IEEEmembership{Fellow,~IEEE}
% }\\
% \thanks{School of Electrical and Information Engineering, University of Sydney, Sydney, NSW 2006, Australia
% E-mail:~\{haiyao.yu,~changyang.she,~yunkai.hu,~geng.wang,~branka.vucetic,~yonghui.li\}@sydney.edu.au}\\
% <-this % stops a space
\author{ \IEEEauthorblockN{Haiyao Yu,
Changyang She,~\IEEEmembership{Senior Member,~IEEE},
Yunkai Hu,
Geng Wang,
Rui Wang, \\
Branka Vucetic,~\IEEEmembership{Life Fellow,~IEEE},
and Yonghui Li,~\IEEEmembership{Fellow,~IEEE}}
\thanks{H. Yu, C. She, Y. Hu, G. Wang, R. Wang, B. Vucetic, and Y. Li are with the School of Electrical and Computer Engineering, University of Sydney, Sydney, NSW 2006, Australia.}

\thanks{E-mails: \{haiyao.yu, ~changyang.she, ~yunkai.hu, ~geng.wang, ~r.wang, ~branka.vucetic, ~yonghui.li\}@sydney.edu.au}
\thanks{The work of Changyang She was supported by the Australian Research Council (ARC) Discovery Early Career Researcher Award (DECRA) under Grant DE210100415. The work of Branka Vucetic was supported in part by the Australian Research Council Laureate Fellowship grant number FL160100032 and the Discovery Project grant number DP210103410. }

}

\maketitle

\begin{abstract}
Machine learning has been considered a promising approach for indoor localization. Nevertheless, the sample efficiency, scalability, and generalization ability remain open issues of implementing learning-based algorithms in practical systems. In this paper, we establish a zero-shot learning framework that does not need real-world measurements in a new communication environment. Specifically, a graph neural network that is scalable to the number of access points (APs) and mobile devices (MDs) is used for obtaining coarse locations of MDs. Based on the coarse locations, the floor-plan image between an MD and an AP is exploited to improve localization accuracy in a floor-plan-aided deep neural network. To further improve the generalization ability, we develop a synthetic data generator that provides synthetic data samples in different scenarios, where real-world samples are not available. We implement the framework in a prototype that estimates the locations of MDs.
Experimental results show that our zero-shot learning method can reduce localization errors by around $30$\% to $55$\% compared with three baselines from the existing literature.
% existing ranging-based methods achieve poor localization accuracy, especially when there is no direct line-of-sight (LoS) path between a mobile device and an access point. To address this issue, We first develop a graph neural network (GNN) that enables multi-device localization by leveraging the round trip time (RTT) and receive signal strength (RSS). The accuracy of estimated locations of MDs is further improved by a novel , which combines the RTT, RSS, and floor plan information. Unlike fingerprint-based methods that require time-consuming and labor-intensive construction of the feature map, our proposed algorithm can be easily transferred to an unseen scenario without additional training samples. We develop a prototype and deploy the proposed algorithm in three different environments with mixed LoS/nLoS conditions. The evaluation results show that in terms of localization accuracy, the proposed algorithm significantly outperforms
% the existing baseline methods, including the hybrid RTT-RSS model, and the Bayesian grid update method. In addition, the proposed algorithm can estimate the locations of MDs that do not have stable connections with three or more APs by RTT distances and RSS between MDs.

%In an office environment with mixed LoS and nLoS paths, the median and $90$-th percentile of the localization errors of our method are 0.7 m and 1.2 m, respectively, compared to 1.56 m and 3.34 m of the existing hybrid RTT-RSS algorithm. 

\end{abstract}

\section{Introduction}

\IEEEPARstart{T}{he} widespread adoption of the global navigation satellite system (GNSS) has revolutionized localization and navigation services, enabling the development of various applications in vertical industries and everyday life \cite{enge1994global}. Although the GNSS can provide real-time localization with meter-level accuracy in most outdoor environments, it is not applicable in indoor environments due to signal blockage \cite{ozsoy2013indoor}. Nevertheless, there has been an increasing demand for indoor localization in different application scenarios, such as smart cities, intelligent plants, and Internet of Things (IoT) \cite{zhu2020indoor}. As an example, mobile users can quickly navigate to a specific store and get to the desired location with the indoor localization system. 

The existing technologies for indoor localization \cite{zafari2019survey} include wireless fidelity (WiFi) \cite{he2015wi}, Bluetooth \cite{jianyong2014rssi}, radio frequency identification device (RFID) \cite{saab2010standalone}, visible light \cite{zhang2016litell}, computer vision \cite{werner2011indoor}, magnetic field (MF) \cite{8292759}, pedestrian dead reckoning (PDR) \cite{8115912}, light detection and ranging (Lidar) \cite{8647251}, and ultra-wideband (UWB) \cite{li2023cooperative}. Among them, visible light and Lidar can achieve the highest positioning accuracy (centimeter-level) in scenarios with line-of-sight (LoS) paths. This fact limits its application in indoor localization, where LoS paths may not be available. The computer-vision-based techniques compare the captured images with pre-stored images to identify the location of the mobile device (MD) \cite{10.1145/3330180.3330193}. This approach is labor intensive as it needs image databases, and it raises potential issues on privacy. Both MF and PDR utilize the micro-electro-mechanical system (MEMS) technologies \cite{glanzer2009semi} that suffer from the accumulated errors \cite{9378557}. For UWB and RFID, we need additional hardware to process the signals, and they are supported by limited cell phones \cite{S24,Pixel}.

The primary demand for an indoor localization system is to choose a technology that is readily accessible at the user end. WiFi has stood out among these technologies due to its wide deployment in indoor environments. In particular, most electronic devices are WiFi-enabled, which makes WiFi an ideal candidate for indoor localization. Depending on the information available in WiFi systems, there are three types of information that can be used for indoor localization: 1) received signal strength (RSS)-based methods, 2)  
channel state information (CSI)-based methods, 3) and round trip time (RTT)-based methods.
%Based on the WiFi standards, MDs and APs can estimate received signal strength (RSS) and channel state information (CSI) for localization. However, existing approaches exhibited performance limitations and drawbacks \cite{tian2016performance,wen2015fundamental,jiokeng2020ftm}. For instance, the distance estimated by the RSS can be seriously affected by signal attenuation from traveling through walls and other obstacles, whereas CSI is difficult to obtain in commercial hardware. \red{Why? Can we address this issue?}

From theoretical path-loss models, RSS can be converted into the propagation distance from an access point (AP) to an MD. Then, ranging-based methods, such as trilateration and least-square algorithm, are applied to estimate the location of the MD \cite{zafari2019survey}. Due to the multi-path propagation of the signals, the accuracy of this approach highly depends on the propagation environment.

% the major challenge associated with this approach is to precisely estimate the signal propagation distance between an MD and an AP. 

CSI is widely used in fingerprint-based localization algorithms. With this approach, we first need to collect the CSI of an MD in the considered area to construct an offline fingerprint data set. For real-time localization, the measured CSI of the MD is matched with the most similar CSI in the data set \cite{zhu2020indoor,singh2021machine,basri2016survey,zheng2023exploiting}. Nevertheless, the fingerprint-based approach heavily relies on offline measurements and is highly sensitive to environmental changes.

In the IEEE 802.11-REVmc2, RTT protocol is available for estimating distances between MDs and APs \cite{ibrahim2018verification}. The RTT protocol does not require clock synchronization between the MD and APs, and is applicable in dynamic communication environments with different types of devices \cite{gentner2020wifi, 9151400}. Nevertheless, in complex indoor environments with non-line-of-sight (nLoS) conditions, the RTT distance could be inaccurate due to multi-path propagation \cite{martin2020ranging}. 

Although we can combine RTT, RSS, or CSI in localization \cite{8924707,dvorecki2019machine}, these types of information may not be sufficient to achieve high localization accuracy. Other types of information are needed, especially in scenarios without LoS paths. For example, 
the results in \cite{wang2020nlos} the geometric relationship among multiple clusters of users can be exploited to mitigate the nLoS effect. In \cite{huang2020hpips}, the environment information is integrated into the localization algorithm. Inspired by these works, we investigate how to exploit floor-plan images in localization algorithms. It's worth noting that floor-plan images can be readily obtained in most indoor scenarios, such as evacuation diagrams. The widespread use and mandatory updating of evacuation diagrams in most buildings ensure accessibility and up-to-date information of the floor plan. Please note that to estimate the location of an MD on a map, which is the floor-plan image in our work, we need to obtain the locations of APs through some simple measurements, which can be done very quickly.

\subsection{Related Work}
%Wi-Fi RTT localization has attracted increasing attention in recent years. Various approaches and methodologies have been proposed to compensate for the RTT ranging error to improve the localization accuracy. 

In LoS environments, the authors in \cite{9151400} developed a clustering-based trilateration algorithm with the weighted concentric circle generation method, which demonstrates good tolerance and positioning accuracy. In \cite{9293264}, the authors utilized the weighted centroid and the particle swarm optimization algorithm to achieve a sub-meter positioning accuracy. 

In mixed LoS/nLoS environments, most existing works first identify LoS and nLoS paths, and then calibrate the ranging errors. For example, in \cite{si2020wi} and \cite{cao2020indoor}, Gaussian mixture models were established to identify nLoS RTT distances. Then, they are removed from the final estimation, and only LoS RTT distances will be preserved for localization. The authors in \cite{han2019smartphone} applied the support vector machine (SVM) in the LoS/nLoS path classification, dividing the signals into high and low-quality ones. In \cite{bregar2018improving}, the authors proposed a convolutional neural network (CNN)-based LoS/nLoS path classification algorithm. In nLoS scenarios, another CNN-based regression algorithm was developed to estimate ranging errors. The authors of \cite{dong2021real} further demonstrated that random forest surpasses SVM and deep neural network (DNN) in terms of the LoS/nLoS classification accuracy. With the help of the classification algorithms in \cite{han2019smartphone,bregar2018improving,dong2021real}, it is possible to achieve higher localization accuracy by removing nLoS paths from localization. Nevertheless, in the scenarios without a strong LoS path, improving the localization accuracy remains a challenging objective.

Deep learning has shown great potential in indoor localization. In \cite{choi2022enhanced}, an unsupervised neural network was developed to improve the RTT ranging accuracy by learning the RTT patterns in different propagation environments. From the signals propagated in different mediums (e.g., metal, wood, etc.) and environments, a deep learning-based framework was proposed in \cite{li2023variational} for estimating the propagation distance. The authors of \cite{seong2021high} designed a hybrid CNN-based compensation distance network with a recurrent neural network (RNN)-based region proposal network to improve the accuracy of a fingerprint-based method. In scenarios with other types of sensors, the authors of \cite{guo2023factor} and \cite{zehua2023indoor} applied a factor graph to integrate different types of information in indoor localization or navigation.

% to limit the
% measurements taken from multiple sensors in the smartphone to a reasonable range and then integrate those measurements to perform localization. 

It is worth noting that most of the existing methods rely on labor-intensive and time-consuming measurements. Although with different training and testing data samples, the samples are collected in a single environment. Improving the generalization ability of learning-based localization algorithms in unseen environments with no or limited measurements deserves further study.

% Most methods mentioned above require  of offline measurements or evaluating the result in a single scenario. The above observations have spurred our interest in advancing WiFi-empowered indoor localization to improve its robustness and achievable localization accuracy.

\subsection{Contributions}

This paper develops a floor-plan-aided indoor localization system that can achieve sub-meter-level accuracy in different indoor environments with mixed LoS and nLoS paths. The main contributions of this paper are summarized as follows: 
%Unlike fingerprint-based methods that require time-consuming and labor-intensive construction of the feature map, our proposed system can be easily deployed and is able to perform accurate localization. 

\begin{itemize}

\item  We establish a multi-user framework that consists of a graph neural network (GNN) and a floor-plan-added deep neural network (FPDNN). By leveraging RTT distances and RSS, the GNN outputs the coarse locations of MDs. Then, the positioning accuracy is further improved by FPDNN, which utilizes the floor-plan image to acquire the signal propagation information in scenarios with mixed LoS and nLoS paths. With the RTT distances and RSS between MDs, our approach can estimate the locations of MDs that do not have stable connections with three or more APs.

\item  To improve the sample efficiency in a real-world environment, we build a virtual environment for synthetic data generation. With the help of synthetic data, the GNN and FPDNN are first pre-trained in the virtual environment and then deployed in the real-world environment. To apply the GNN and FPDNN in an unseen real-world environment with no data samples, we only need the floor-plan image to update the virtual environment. This can help to reduce labor and time consumption for measurement in the new environment.  If some data samples are available, we can further increase the accuracy of our localization algorithm by fine-tuning the GNN and FPDNN in the new environment.

%synthetic data generator to increase the generalization ability, which can assist in transferring our proposed system into different environments without additional collected samples. 

%to further enhance the localization accuracy, we proposed a novel FPDNN model to minimize the RTT ranging error by learning the relation between RTT distance, RSS, and spatial information derived from the floor plan. Especially in the scenarios without LoS path, our proposed model can improve the RTT distance by learning the environmental structure (e.g., the number of walls, their thickness and materials) from the floor plan. 

\item To collect data samples and to evaluate the proposed localization algorithm, we develop a prototype and deploy it in three different environments: office, laboratory, and shopping mall. We collect around $100,000$ samples in total. In each environment, the RTT distances and RSS from the MDs to all the APs are updated to our server in real-time, and the locations of MDs can be updated every $200$~ms.

%The first two data sets include  and the inertial measurement unit (IMU) of the MDs. In addition to this information, the third data set provides the . For each data set, we collect data samples from three scenarios. 

%\item Finally, we deployed our system on commodity devices and evaluated the localization performance based on field experiments. The experimental results prove that the proposed model can achieve a sub-meter-level localization accuracy in the environment with mixed LoS and nLoS paths. Furthermore, we evaluate the robustness of the proposed model by deploying the model in different environments. Specifically, we generate the synthetic data in the virtual environment to train the proposed model and evaluate it in the real-world scenario. The experiment result indicates that our proposed model can be easily deployed to new environments and the accuracy can be further improved with only a small amount of labeled data collected by transfer learning.

\item Our field experiments show that the proposed framework can achieve sub-meter-level accuracy in environments with mixed LoS and nLoS conditions. Specifically, the root mean square error (RMSE) in the office, laboratory, and shopping mall are $0.87$~m,  $1.14$~m, and $0.9$~m, respectively. With our zero-shot learning approaches, the RMSE is around $30$\% $\sim 55$\% lower than the baselines in \cite{8924707}, \cite{horn2022indoor} and \cite{9151400}.
\end{itemize}

\subsection{Notations}

In this paper, we use uppercase letters, e.g., $X$, to represent given constant numbers or parameters. Lowercase letters, e.g., $x$, are used to represent scalar variables. We use a calligraphic uppercase letter, e.g., $\mathcal{A}$, to denote a set, and $|\mathcal{A}|$ is its cardinality. In particular, $\{x_{k}\}_{k = 1}^{K}$ denotes a set with given elements, i.e., $\{x_{k}\}_{k = 1}^{K} = \{x_{1},\ldots,x_{K}\}$, and $\mathbb{R}$ is the set of real numbers. We use an uppercase bold letter, e.g., $\mathbf A$, to represent a matrix, and a lowercase bold letter, e.g., $\mathbf{a}$, to denote a row vector. Superscript $^\top$ denotes the transposition of a matrix or vector, e.g., $\mathbf A^\top$ and $\mathbf a^\top$, respectively. For matrices $\mathbf{A}$ and $\mathbf{B}$, we define $[\mathbf{A},\mathbf{B}]$ as their horizontal concatenation, while $\begin{bmatrix}
  \mathbf{A} \\
  \mathbf{B}
\end{bmatrix} = [\mathbf{A}^\top,\mathbf{B}^\top]^\top$ is the vertical concatenation. Other notations will be specifically stated.

\begin{figure}[tbp]

	\vspace{-0.1cm}
	
	\begin{minipage}[t]{0.5\textwidth}
		\centering\includegraphics[width=0.8\textwidth]{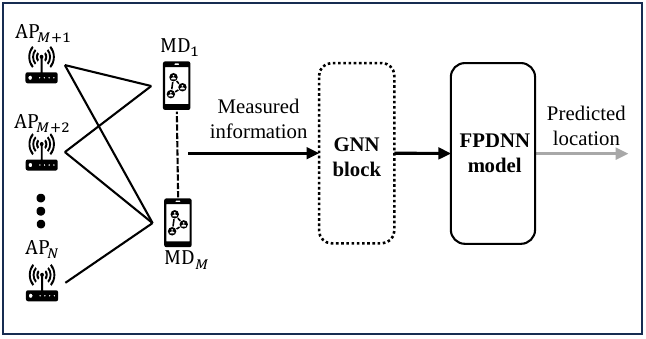}
	\end{minipage}
 
	%\vspace{-0.1cm}
	\caption{Framework of the localization system.}
	\label{fig:systemmodelmultiuser}
	\vspace{-0.3cm}
\end{figure}

\section{Framework of Localization System}\label{image model}

In this section, we elaborate on the framework of our proposed localization system. As depicted in Fig.~\ref{fig:systemmodelmultiuser}, there are two blocks in the proposed framework: 1) the graph-based pre-localization, and 2) FPDNN for improving the localization accuracy. Details of these blocks will be elaborated in subsequent sections.

 %There are two different scenarios in the system: the real-world environment and the virtual environment. The offline stage, including the graph-based pre-localization and the synthetic data generation, is trained by the data samples collected in the virtual environments, and the online training stage is trained by the data collected in the real-world environment.

%\subsection{WiFi Platform}
%We note that the synthetic data generation and the graph-based pre-localization stage are offline stages, which are pre-trained in the virtual environment, while the online fine-tuning stage is trained in the real-world environment.

We consider a localization system with $M$ MDs and $K$  WiFi APs. Thus, there are a total of $N=M+K$ nodes in the network. The indices of MDs are denoted by $m \in \mathcal{M}$,  where $\mathcal{M} \triangleq \{1,\ldots,M\}$, and the indices of APs are denoted by $k \in \mathcal{K}$, where $\mathcal{K} \triangleq \{M+1,\ldots,M+K\}$. In some cases where we do not need to distinguish MDs and APs, we will use index $n$ to refer to the $n$-th node. 

%In order to distinguish the difference between the two environments, the indices with superscript $'$, e.g., $m',k'$ denote the indices in the virtual environment, and the $m, k$ denote the indices in the real-world environment. For example, the locations of the MD and the APs at the $t$-th time slot are denoted by $\Upsilon_{m'}^{(t)}$ and $\Upsilon_{k'}$ respectively in the virtual environment, and $\Upsilon_{m}^{(t)}$ and $\Upsilon_k$ in the real-world environment.

Time is discretized into time intervals. In each time interval, an MD scans the nearby nodes (including both APs and MDs) and broadcasts the initial RTT request. The nodes respond to the request and send the acknowledgment back to the MD to establish the connection. Once the connection is established, the MD starts measuring the distance to nearby nodes, including MDs and APs \cite{martin2020ranging}. The measurement between the $m$-th MD and the $n$-th node includes their RTT distance, $\zeta^{(t)}_{m,n}$, RSS, $\gamma^{(t)}_{m,n}$, and the index of the node. Thus, the information obtained by the $m$-th MD is given by
\begin{equation}
{\bm p}^{(t)}_{m,n}=[\zeta^{(t)}_{m,n}, \gamma^{(t)}_{m,n},n], n \ne m.
\end{equation}
Since an MD cannot establish a link to itself, we define ${\bm p}^{(t)}_{m,m} =\emptyset $.

%where  is  between the $m$-th MD and the $n$-th node, $\sigma^{(t)}_{m,n}$ 
% The FIL mainly includes three parts: FPDNN, the graph construction block, and the GRNN. 

In the $t$-th time interval, all the MDs upload their information to a server. The information that can be used for localization is denoted by $\mathbf{P}^{(t)}$, where
\begin{equation}
\mathbf{P}^{(t)}=
  \begin{bmatrix}
    {\bm p}^{(t)}_{1,1}  & {\bm p}^{(t)}_{2,1}& \dots &{\bm p}^{(t)}_{m,1}  & \dots  & {\bm p}^{(t)}_{M,1} \\
    {\bm p}^{(t)}_{1,2} &  {\bm p}^{(t)}_{2,2} & \dots  &{\bm p}^{(t)}_{m,2} & \dots  & {\bm p}^{(t)}_{M,2} \\
    \vdots &  \vdots & \dots& \vdots & \ddots & \vdots \\
    {\bm p}^{(t)}_{1,N}  & {\bm p}^{(t)}_{2,N}& \dots & {\bm p}^{(t)}_{m,N}&\dots  & {\bm p}^{(t)}_{M,N}
\end{bmatrix}. 
\end{equation}

From $\mathbf{P}^{(t)}$, the GNN estimates the coarse locations of MDs. Based on the coarse locations, the floor-plan image from each AP to the $m$-th MD is cropped from the floor-plan image, which is exploited by FPDNN to further improve localization accuracy. We note that the coarse locations obtained from the GNN help FPDNN to identify the communication environment between the MD and its neighbors. Thus, a better pre-localization algorithm helps to improve the final localization accuracy.

\section {GNN Pre-localization}
In this section, we first introduce how to construct a graph representation of a wireless network. Then, we develop a GNN for pre-localization to get the coarse locations of MDs.

\subsection{Graph Representation}\label{graphconstructionblock}

\begin{figure}[htp]
	\vspace{-0.1cm}
	\centering
	\begin{minipage}[t]{0.5\textwidth}
		\includegraphics[width=1.0\textwidth]{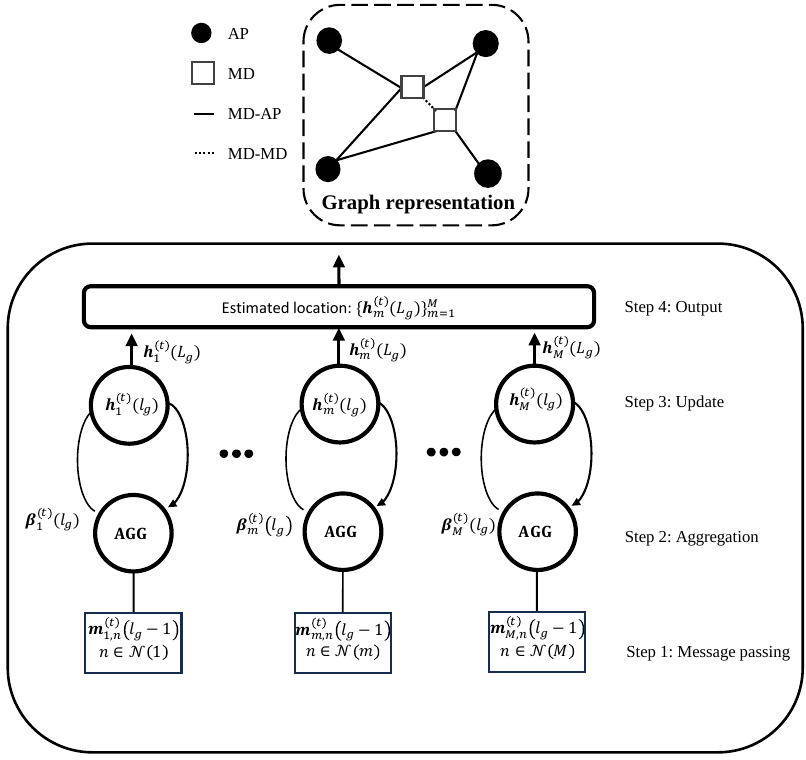}
	\end{minipage}
	%\vspace{-0.1cm}
	\caption{Structure of GNN.}
	\label{fig:grnn}
	%\vspace{-0.3cm}
\end{figure}
We denote the graph representation at the $t$-th time interval by ${\mathcal{G}}^{(t)}$. As depicted in Fig.~\ref{fig:grnn}, the circles are the APs, and the squares are the MDs. Each solid line is an edge between an MD and an AP, and the dashed line represents the edge between two MDs. The node and edge features are summarized as follows: 
% The line between an MD and an AP is the edge between them. The node and edge features are summarized as follows:

\begin{itemize}

\item The node feature of an AP is the coordinate of it,
\begin{equation*}
{\Upsilon}_k=({x}_k,{y}_k), {k=M+1,M+2,\ldots ,N},    
\end{equation*}
which remain constant in the localization system.
\item The node feature of an MD includes the estimated location of it. The estimated location is updated by a GNN with $L_g$ layers. We denote the edge feature in the $l_g$-th layer of the GNN by ${\bm h}_{m}^{(t)}(l_g) \in \mathbb{R}^{2 \times 1}$. The initial node feature ${\bm h}_{m}^{(t)}(0)$ is the estimated location obtained from the trilateration method that only exploits the RTT distances, ${\zeta}^{(t)}_{m,n}$.

% \begin{equation*}
% {{\Upsilon}^{\mathrm{Tri}}_m}^{(t)}=f^\mathrm{Tri}(\{{\zeta}^{(t)}_{m,k}; {\Upsilon}_k\}_{k \in \mathcal{K}_m^{(t)}}).  
% \end{equation*}
%For the MDs that cannot establish connections with at least three APs, we set ${\hat{\Upsilon}}^{(t)}_m=(0,0)$. \blue{(Am I right?)}

\item Let $\mathcal{K}_m^{(t)}$ and $\mathcal{M}_m^{(t)}$ be the sets of APs and MDs that can establish stable connections with the $m$-th MD at the $t$-th time interval, respectively. The set of neighbor nodes of the $m$-th MD is defined as $\mathcal{N}_m \triangleq \mathcal{K}_m^{(t)} \cup  \mathcal{M}_m^{(t)}$. The edge feature between the $m$-th MD and the $n$-th node is defined as
\begin{equation*}
{\bm e}^{(t)}_{m,n}=[{\zeta}^{(t)}_{m,n},{\gamma}^{(t)}_{m,n},n],\forall {n \in \mathcal{N}_m}.
\end{equation*}
% where $\mathcal{K}_m^{(t)}$ is the set of APs, in which the distance between the MD and AP is less than $25$~m.

% \item The edge feature between each MD-MD pair is 
% \begin{equation*}
% {\bm e}^{(t)}_{m,\hat{m}}=[\zeta^{(t)}_{m,\hat{m}}, \gamma^{(t)}_{m,\hat{m}},\hat{m}],
% \end{equation*}
% where $\hat{m}\in \mathcal{M}$ and $\hat{m}\neq m$.
\end{itemize} 
We assume that two MDs can establish a stable connection between them when there is a LoS path. In nLoS cases, there is no edge between two MDs.

%It is important to note that since the measurements between two MDs are unstable in the nLoS scenario, we only construct the edge between MDs in the LoS scenario. This condition can be distinguished simply by setting the threshold for the standard deviation of the measurement.

\subsection{Graph Neural Network}\label{GNN block}

The GNN is employed for the pre-localization of MDs. The input of the GNN block is the initial graph representation at the $t$-th time interval, ${\mathcal{G}}^{(t)}$. The output of the GNN block is the coarse locations of MDs, ${\Tilde{\Upsilon}}^{(t)}_m=(\Tilde{x}_m,\Tilde{y}_m)^{(t)}, m=1,2,\ldots M$.

As shown in Fig.~\ref{fig:grnn}, the GNN consists of four steps: 1)  message-passing, 2)  aggregation, 3) feature update, and 4) output.

\begin{enumerate}
\item Message-passing: The GNN consists of $L_g$ layers. The input of the $m$-th MD is ${\bm h}^{(t)}_{m}(0)$, which is also the initial feature of the MD. In the $l_g$-th layer of the GNN, to update the feature of the $m$-th MD, all its neighbor nodes generate messages based on their features, ${\bm h}^{(t)}_{n}(l_g-1)$, and the edge feature between the $m$-th MD and the $n$-th node, ${\bm e}^{(t)}_{m,n}$, $n \in \mathcal{N}_m$. The message can be expressed as
\begin{equation}
{\bm m}^{(t)}_{m,n}(l_g)=\phi\Big({\bm h}^{(t)}_{n}(l_g-1),{\bm e}^{(t)}_{m,n};{\bm \theta}_{\rm{M}}\Big),
\end{equation}
where $\phi({\cdot};{\bm \theta}_{\rm{M}})$ represents an feed forward neural network (FFNN) with parameters ${\bm \theta}_{\rm{M}}$.

\item Aggregation: The $m$-th MD aggregates the messages from all its neighbors to obtain aggregated information $\bm{\beta}^{(t)}_m(l_g)$. The aggregation can be expressed as
\begin{equation}
\bm{\beta}^{(t)}_m(l_g)={\rm AGG}\Big({\bm m}^{(t)}_{m,n}(l_g), n\in \mathcal{N}_m\Big).\end{equation}
In our GNN, we set the aggregation function to the mean of all the messages.

\item Feature update: The $m$-th MD updates its feature by merging ${\bm h}^{(t)}_{m}(l_g-1)$ and $\bm{\beta}^{(t)}_m(l_g)$. The output of the $l_g$-th layer can be expressed as 
\begin{equation}
{\bm h}^{(t)}_{m}(l_g)=\varphi \Big({\bm h}^{(t)}_{m}(l_g-1),\bm{\beta}^{(t)}_m(l_g);{\bm \theta}_{\rm{F}}\Big),
\end{equation}
where ${\varphi}({\cdot};{\bm \theta}_{\rm{F}})$ is an FFNN with parameters ${\bm \theta}_{\rm{F}}$.

\item Output: After $L_g$ rounds of updates, we can obtain the output from all the nodes, $\{{\bm h}^{(t)}_{m}(L_g)\}_{m=1}^M$. The output of the GNN is the estimated locations of all the MDs, ${\Tilde{\Upsilon}}^{(t)}_m = {\bm h}^{(t)}_{m}(L_g), m = 1,2,..., M$.
\end{enumerate}

To train the GNN, we use the ground truth locations of MDs as labels, and minimize the gap between labels and the outputs of the GNN, e.g.,
\begin{align}\label{eq:LossGNN}
\mathcal{L}_{\rm GNN}({\bm \theta}_{\rm{M}},{\bm \theta}_{\rm{F}}) = {\mathbb{E}}(||{\Tilde{\Upsilon}}^{(t)}_m - {{\Upsilon}}^{(t)}_m||^2),
\end{align}
where ${{\Upsilon}}^{(t)}_m \triangleq (x_m^{(t)},y_m^{(t)})$ is the ground truth of the location, and $||{\Tilde{\Upsilon}}^{(t)}_m - {{\Upsilon}}^{(t)}_m||^2$ is the Euclidean distance between ${\Tilde{\Upsilon}}^{(t)}_m$ and ${{\Upsilon}}^{(t)}_m$. With the loss function in Eq.~\eqref{eq:LossGNN}, we can use stochastic gradient descent to train the parameters ${\bm \theta}_{\rm{M}}$ and ${\bm \theta}_{\rm{F}}$ of the GNN.
 
\begin{figure}[htp]
	%\vspace{-0.1cm}
	\centering
	\begin{minipage}[t]{0.49\textwidth}
		\includegraphics[width=1.0\textwidth]{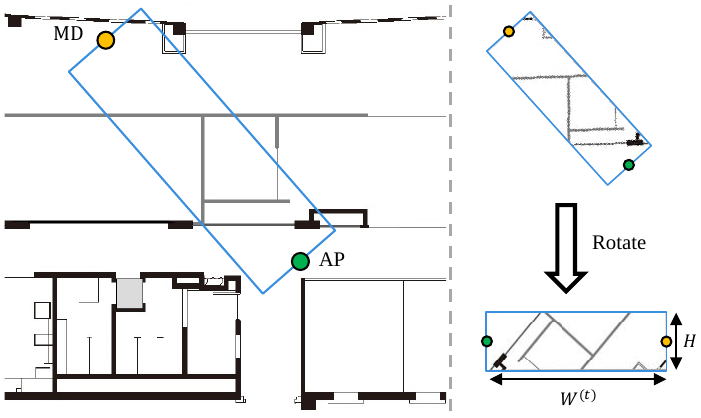}
	\end{minipage}
	%\vspace{-0.1cm}
	\caption{Image preparation.}
	\label{fig:imageconstruct}
	\vspace{-0.3cm}
\end{figure}

\begin{figure*}[tbp]
	\centering
	\subfigure[FPDNN for estimating distance between the $k$-th AP and the $m$-th MD. ]{
		\begin{minipage}[b]{0.8\textwidth}
			\includegraphics[width=1\textwidth]{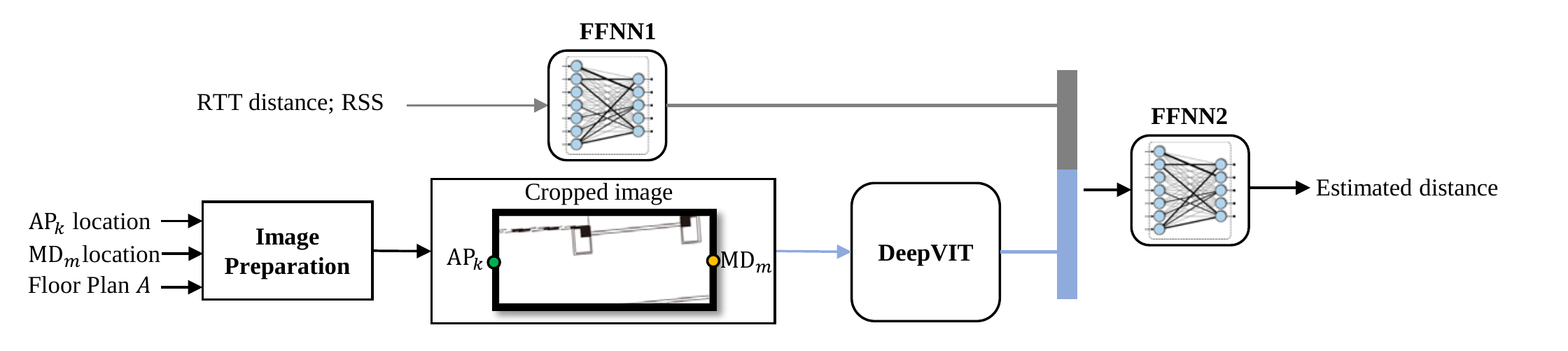} 
		\end{minipage}
	\label{fig:FPDNN}
	}
    	\subfigure[Synthetic data generator]{
    		\begin{minipage}[b]{0.8\textwidth}
   		 	\includegraphics[width=1\textwidth]{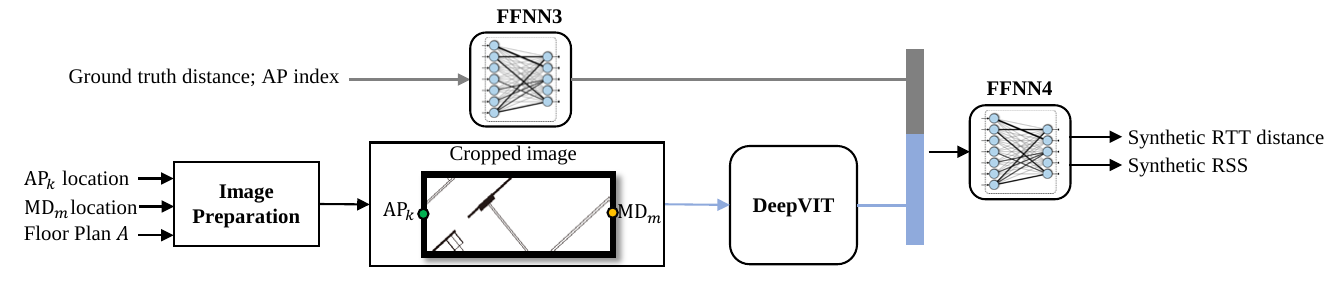}
    		\end{minipage}
		\label{fig:synetic_data_generator}
    	}    
	\caption{Structure of FPDNN and synthetic data generator.}
 \vspace{-0.3cm}
	\label{fig:FPDNN_data_generator}
\end{figure*}

% \begin{figure*}[htbp]
% 	\vspace{-0cm}
% 	\centering
% \hspace{0.3cm}
% 	\begin{minipage}[t]{0.8\textwidth}
% 		\includegraphics[width=1\textwidth]{draft/FPDNN_model.pdf}
% 	\end{minipage}
% 	%\vspace{-0.1cm}
% 	\caption{Illustration of the FPDNN model for the pair of $k$-th AP and $m$-th MD.}
% 	\label{fig:FPDNN}
% 	\vspace{-0.3cm}
% \end{figure*}

\section {Floor-plan-aided Deep Neural Network}
In this section, we introduce FPDNN for a better estimation of the distance between an MD and an AP.

%The FPDNN model further increases the accuracy of the coarse location based on the floor plan image. Specifically, when an MD connects to at least $3$ APs, the image generator will crop the image between the coarse location of the MD and each connected AP. Then the cropped images and the calibrated RTT distance, RSS are fed as the input to the FPDNN, which helps extract the information of the propagation scenario. The output of FPDNN is the predicted distance between each MD and AP pair. Finally, these distances are converted to the location of $m$-th MD based on the least square algorithm. Details of the image preparation and the FPDNN structure will be elaborated in subsequent sections.

%In this section, we first introduce the image generator, which generates the cropped image between each MD-AP pair as the input to FPDNN. Then, we illustrate the structure of FPDNN.
\subsection{Image Preparation}
%\subsubsection{Floor plan preparation}
As we mentioned, the RTT distance is biased due to the multi-path effect, especially when the LoS path is blocked. To improve localization accuracy, we utilize the floor-plan image between an MD and an AP. The images are cropped from the floor plan of the all area.

%Thus, the multi-path and delay information is extracted from the position of the blockage and the corresponding thickness. In order to assist the FPDNN in acquiring the information, the floor plan is needed. As shown in Fig.~\ref{fig:imageconstruct}, the position and the thickness of the blockage is presented by different lines with different width. 

Fig.~\ref{fig:imageconstruct} shows the procedure for preparing images between the $m$-th MD and the $k$-th AP. We first find the coarse location of the MD estimated by the GNN block. Then, we crop a rectangular image with size $\{{W}^{(t)} \times H\}$ between the MD and the AP, where $W^{(t)}$ is the coarse distance between the MD and the AP, and $H$ is the width of the image. It is worth noting that the selection of width $H$ depends on the positioning accuracy of the coarse location. Specifically, the width $H$ of the cropped image is set to 256, corresponding to approximately $4$~m in the floor-plan image. The RMSE of the coarse location estimated by the GNN block is less than $2$~m. Thus, the redundancy is sufficient for extracting geometric information surrounding the MD.

To regulate the cropped image, we rotate the image such that the direction from the AP to the MD in the image is always pointing rightward, as shown in Fig.~\ref{fig:imageconstruct}. We denote the image between the $m$-th MD and the $k$-th AP by ${{{\mathbf I}}}_{m,k}^{(t)} \in \mathbb{R}^{W^{(t)}\times H }$, which can be obtained from the following function,%\blue{(Is it one image or a set of images? Be careful with your notations? What is the subscript? Does it depend on $t$? It is not consistent with the notations in the sequel.)} %with a size of $\{W\times H\}$.  
\begin{equation}
\begin{aligned}
&{\mathbf{I}}_{m,k}^{(t)}= f_{\mathrm{IG}}({\Tilde{\Upsilon}}_m^{(t)},{{\Upsilon}}_k,\mathbf{A}),
\end{aligned}
\end{equation}
where $\mathbf{A}$ is the floor plan of the whole area, and $f_{\mathrm{IG}}(\cdot,\cdot,\cdot)$ includes the image cropping and rotation.

Therefore, the input features of FPDNN include 1) the collected information from the MD, $\mathbf{P}_m^{(t)}$; 2) the images between the MD and its neighbor APs, $\{{\mathbf{I}}^{(t)}_{m,k}\} \forall{{k\in\mathcal{K}_m^{(t)}} }$.

%We note that for the image generator in the virtual environment, the output is denoted as ${\mathbf{I}}_{m',k'}$.

 %The structure of these blocks will introduced in the following subsections.%We note that the structure of these blocks has a similar structure to the DNN model introduced in Section.~\ref{sec:deepvit} and Section.~\ref{sec:FNN layer}. Specifically, all FNNs have the ReLu activation function Eq.~\eqref{eq:relu} and the hidden layer output form Eq.~\eqref{eq:FNN}, while the dimensions of hidden layers can be different (see Table.~\ref{tab:parameter} of Section~\ref{numerical}). We note that the deepVIT has the same structure introduced in Section.~\ref{sec:deepvit}.

\subsection{Structure of FPDNN}

In Fig.~\ref{fig:FPDNN}, we illustrate the structure of FPDNN and show how to estimate the distance between the $m$-th MD and the $k$-AP. FPDNN consists of three blocks: the deep vision transformer (DeepVIT) block and two FFNN blocks. At the $t$-th time interval, the image feature ${\mathbf{I}}^{(t)}_{m,k}$ and the measurement 
$\mathbf{p}_{m,k}^{(t)}$ are the inputs of the DeepVIT block and the first FFNN, respectively. The concatenation of their outputs serves as the input of the second FFNN, which outputs the estimated distance, denoted $\hat{\zeta}^{(t)}_{m,k}$. The true distance between the AP and the MD, denoted by $d_{m,k}^{(t)}$, is used as the label to train FPDNN.

The structures of the DeepVIT block and the two FFNNs will be further elaborated in the sequel.
\subsubsection{DeepVIT block}\label{sec:deepvit}
\begin{figure}[h]
	%\vspace{-0.1cm}
	\centering
	\begin{minipage}[t]{0.47\textwidth}
		\includegraphics[width=1\textwidth]{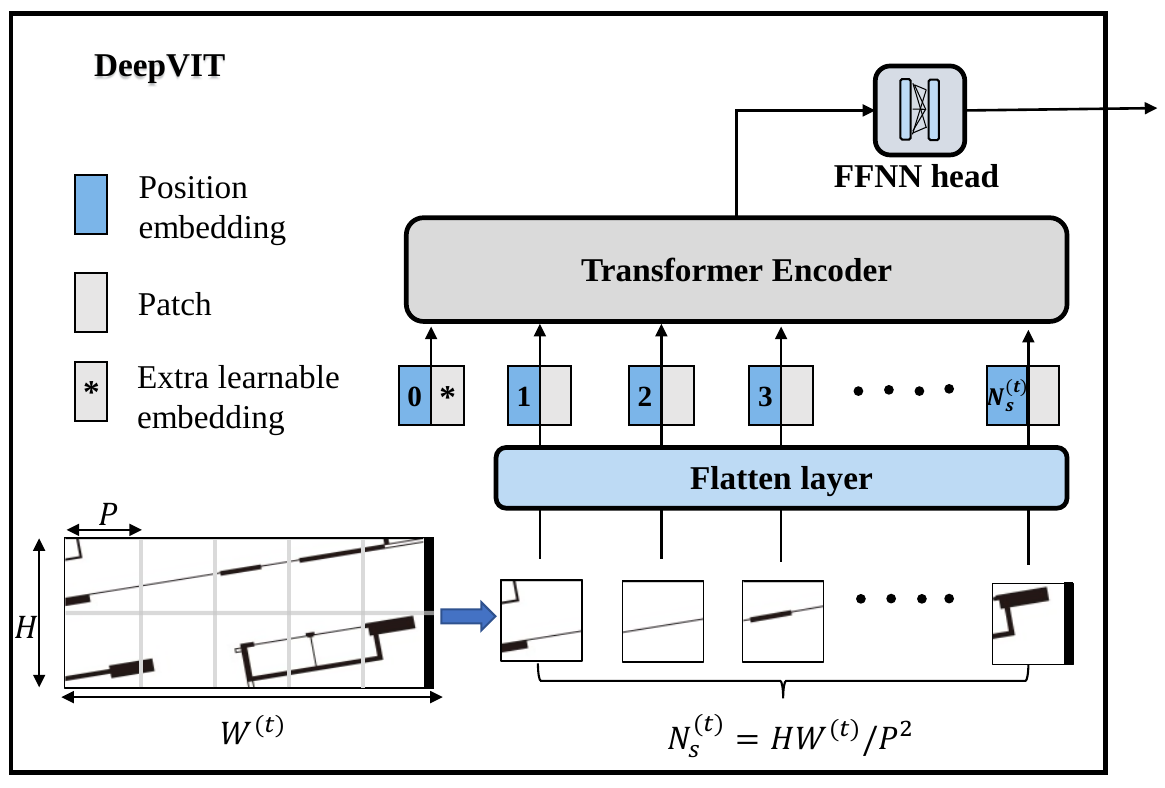}
	\end{minipage}
	%\vspace{-0.1cm}
	\caption{Illustrate the structure of DeepVIT.}
	\label{fig:DeepVIT}
	\vspace{-0.3cm}
\end{figure}

DeepVIT is a kind of transformer that assigns weighting parameters to different parts of the image. The parts with higher weighting parameters will have stronger impacts on the final output. In this way, the output can “pay attention” to the important parts of the image. 

The input of the DeepVIT block is the floor-plan image ${\mathbf{I}}^{(t)}_{m,k}$ and the output of the DeepVIT is the hidden feature given by $\mathbf{o}_{\mathrm{T}}$. There are three different layers in the DeepVIT block, including the splitting and flattening layer, the position embedding layer, and the transformer encoder layer. 
% \begin{equation}\label{eq:transformerenoder}
% \begin{aligned}
% &\mathbf{o}_{\mathrm{T}}= f_{\mathrm{T}}({{\mathbf{I}}^{(t)}_{m,k}};\bm{\theta}_{\mathrm{T}}),
% \end{aligned}
% \end{equation}
% where $f_{\mathrm{T}}$ represents the layers of the deepVIT, and $\bm{\theta}_{\mathrm{T}}$ are the corresponding trainable parameters.

\begin{itemize}
     \item \textit{Splitting and flattening}:
As shown in Fig.~\ref{fig:DeepVIT}, the input image $\mathbf{I}$ is split into $N_{s}^{(t)} =W^{(t)}H/P^{2} $ mini-images. Let ${\mathbf I}_{\ell} \in \mathbb{R}^{P\times P}$ denote the $\ell$-th mini-image, where $P$ is the side length of each mini-page. Then, each mini-image is vectorized to a vector $\mathbf{s}_{\ell}$ with dimension $D'=P\times P$. For instance, let $\mathcal{I}_{i,j,k}$ denote the element of ${{\mathbf I}}_{\ell}$ at the position $(i,j,k)$, and then
\begin{equation}
\begin{aligned}
\mathbf{s}_{\ell} = &[\mathcal{I}_{1,1,1}, \mathcal{I}_{1,1,2},\ldots, \mathcal{I}_{1,1,C},  \mathcal{I}_{1,2,1} \ldots, \mathcal{I}_{1,P,C}, \\ & \mathcal{I}_{2,1,1}, 
\mathcal{I}_{2,1,2},\ldots, \mathcal{I}_{2,P,C}, \mathcal{I}_{3,1,1}, \ldots, \mathcal{I}_{P,P,C}].
\end{aligned}
\end{equation}
Next, each vector $\mathbf{s}_{\ell}$ is compressed into the patch $\mathbf{s}_{\ell}\mathbf{E}$ with the size of $1\times D$, where ${\mathbf E}\in  \mathbb{R}^{D'\times D}$ denotes a trainable matrix. Finally, The output of the flatten layer, denoted as $\mathbf{F}$, is expressed as follows:
\begin{equation}\label{eq:split_flatten}
\begin{aligned}
{\mathbf{F}}= 
\begin{bmatrix}
  \mathbf{s}_{1} {\mathbf E} \\
  \mathbf{s}_{2} {\mathbf E}\\
  \vdots\\
  \mathbf{s}_{N_s}{\mathbf E}
\end{bmatrix},
\end{aligned}
\end{equation}
where $\mathbf{F} \in \mathbb{R}^{N_s^{(t)}\times D}$.

 \item \textit{Position embedding:}
Following the approach in \cite{dosovitskiy2020image}, an extra learnable embedding $\mathbf{s}_{class} \in \mathbb {R}^{1\times D}$ is prepended to $\mathbf{F}$, resulting in $
\begin{bmatrix}
  \mathbf{s}_{class} \\
  {\mathbf{F}}
\end{bmatrix}$. Then, the position embedding, denoted as $\mathbf{E}_{pos}$, is applied to obtain the input of the transformer encoder, ${\mathbf U}_{\mathrm{T}}$. Specifically,
\begin{equation}\label{eq:split_flatten_2}
\begin{aligned}
&{\mathbf U}_{\mathrm{T}}= \begin{bmatrix}
  \mathbf{s}_{class} \\
  {\mathbf{F}}
\end{bmatrix}+{\mathbf E}_{pos},
\end{aligned}
\end{equation}
where ${\mathbf E}_{pos}\in  \mathbb {R}^{(N_{s}^{(t)}+1)\times D}$. Please refer to \cite{dosovitskiy2020image} for more details on position embedding.
 \item \textit{Transformer encoder:}
The transformer encoder applied in this paper has the same structure as \cite{zhou2021deepvit}, which is comprised of two norm layers, a re-attention layer, and an FFNN head. It employs re-attention mechanisms to discern relationships among mini-images, thereby enhancing system performance. Given the input ${\mathbf U}_{\mathrm{T}}$, the output of the transformer encoder can be expressed as
\begin{equation}\label{eq:transformerenoder}
\begin{aligned}
&\mathbf{o}_{\mathrm{T}}= f_{\mathrm{T}}({\mathbf U}_{\mathrm{T}};\bm{\theta}_{\mathrm{T}}),
\end{aligned}
\end{equation}
%\green{check if $\mathbf{v}$ is vector or matrix, I changed $\mathbf{T}$ to $\mathbf{v}$}\red{v is vector}
where $f_{\mathrm{T}}$ is a transformer encoder neural network, and $\bm{\theta}_{\mathrm{T}}$ is the corresponding trainable parameters.

% Following the output of the transformer encoder, there is an MLP head. As introduced in \cite{ba2016layer}, the MLP head consists of a normalization layer and an FFNN. 
 \end{itemize}

\subsubsection{FFNN layers}\label{sec:FFNN layer} 
The collected information $\bm{p}^{(t)}_{m,k}$ serves as input of the FFNN$1$.  The first FFNN takes the measured information, $\bm{p}^{(t)}_{m,k}$, as its input and output some hidden features according to 
\begin{equation}\label{eq:FFNN}
\begin{aligned}
&\mathbf{o}_{\rm F1}=f_{\rm F1 }(\bm{p}^{(t)}_{m,k}, \bm{\theta}_{\rm F1}),
\end{aligned}
\end{equation}
where $\bm{\theta}_{\rm F1}$ are the training parameters.

% ${\color{red} how to put the one in FFNN?}$
% The output from the $\ell$-th hidden layer of the FFNN is denoted by $\mathbf{o}_{FFNN}^{[\ell]} \in \mathbb{R}^{\kappa^{[\ell]}\times 1} $, where $\kappa^{[\ell]}$ is the dimension of the output. ${\color{red}why relation?}$ The relationship between the $\mathbf{o}_{FFNN}^{[\ell-1]}$ and $\mathbf{o}_{FFNN}^{[\ell]}$ can be expressed as
% \begin{equation}\label{eq:FFNN}
% \begin{aligned}
% &\mathbf{o}_{FFNN}^{[\ell]}=f_{r}^{[\ell]}(\mathbf{W}^{[\ell]}\mathbf{o}_{FFNN}^{[\ell-1]}+\mathbf{b}^{[\ell]}),
% \end{aligned}
% \end{equation}
% where $\mathbf{W}^{[\ell]} \in \mathbb{R}^{\kappa^{[\ell]}\times \kappa^{[\ell-1]}}$, $\mathbf{b}^{[\ell]}\in \mathbb{R}^{\kappa^{[\ell]}\times 1}$, and $f_{r}^{[\ell]}$ is the ReLu function
% \begin{equation}\label{eq:relu}
% \begin{aligned}
% &f_{r}^{[\ell]}(x)=\max(x,0).
% \end{aligned}
% \end{equatio

The input of the second FFNN is ${\mathbf u}_{\rm{F2}}=[\mathbf{o}_{\rm{T}},{\mathbf o}_{\rm{F1}}]$. It outputs the estimated distance between the $m$-th MD and the $k$-th AP,
\begin{equation}\label{eq:outFPDNN}
\begin{aligned}
\hat{\zeta}^{(t)}_{m,k}=f_{\rm F2 }\big({\mathbf u}_{\rm{F2}};{\bm \theta_{\rm{F2}}}\big).
\end{aligned}
\end{equation}

%\subsection{Training of FPDNN}
% Given the output of the FPDNN in \eqref{eq:outFPDNN}, ${\mathbf{I}}^{(t)}_{m,k}$ and $\hat{\zeta}^{(t)}_{m,k}$, 

To train FPDNN, we use the ground truth distances between the $m$-th MD and the $k$-th AP ${d}^{(t)}_{m,k}$ as the labels. The loss function is defined as the mean square error (MSE) between the output of FPDNN and the label,
\begin{equation}\label{eq:gradient}
\begin{aligned}
\mathcal{L} ({\mathbf E}, {\bm \theta}_{\mathrm{T}},{\bm \theta}_{\mathrm{F1}},{\bm \theta}_{\mathrm{F2}})= {\mathbb{E}}\bigl(\hat{\zeta}^{(t)}_{m,k}-{d}^{(t)}_{m,k}\bigl)^2.
\end{aligned}
\end{equation}
The parameters of FPDNN are updated by the stochastic gradient descent algorithm \cite{bottou2010large}.

% i.e.,
% \begin{equation}
%     {\bm \theta}_{\mathrm{F}}= {\bm \theta}_{\mathrm{F}}-\eta\frac{\partial \mathcal{L}({\bm \theta}_{\mathrm{F}})}{\partial {\bm \theta}_{\mathrm{F}}},
%     \end{equation}
% where $\eta$ is the learning rate.

% \begin{figure*}[htbp]
% 	\centering
% 	\subfigure[Synthetic RTT.]{
% 		\begin{minipage}[b]{0.45\textwidth}
% 			\includegraphics[width=1\textwidth]{draft/synethic_RTT.jpg} 
% 		\end{minipage}
% 	\label{fig:synetic_rtt}
% 	}
%     	\subfigure[Synthetic RSS.]{
%     		\begin{minipage}[b]{0.45\textwidth}
%    		 	\includegraphics[width=1\textwidth]{draft/synethic_RSS.jpg}
%     		\end{minipage}
% 		\label{fig:synetic_rss}
%     	}
	   
% 	\caption{Synthetic data generator in the office.}
%  \vspace{-0.3cm}
% 	\label{fig:synethic}
% \end{figure*}

\subsection{The Least Square Block}\label{LS_algorithm}
After obtaining the estimated distances from the $m$-th MD to all its neighbor APs, $\{\hat{\zeta}^{(t)}_{m,k}\}_{k \in \mathcal{K}_m^{(t)}}$, the least square algorithm is used to estimate the location of the MD \cite{4212819}. The output of the least square can be expressed as 
\begin{equation}
% \{{\hat{\Upsilon}}^{(t)}_m=(\hat{x}_m,\hat{y}_m)^{(t)}\}^{M}_{m=1}.  
{{\hat{\Upsilon}}^{(t)}_m}=f^\mathrm{LS}(\{\hat{\zeta}^{(t)}_{m,k}, {\Upsilon}_k\}_{k \in \mathcal{K}_m^{(t)}}),
\end{equation}
where $f^\mathrm{LS}(\cdot)$ is the least square algorithm, and ${{\hat{\Upsilon}}^{(t)}_m} \triangleq (\hat{x}_m,\hat{y}_m)^{(t)}$ is the location of the $m$-th MD.

In addition, the Kalman filter could be applied to improve the localization accuracy in the current time interval by using the historical trajectory of the MD.

\section{Synthetic Data Generation in Virtual Environments}
\begin{figure}[htbp]
	%\vspace{-0.1cm}
	\centering
	\begin{minipage}[t]{0.4\textwidth}
		\includegraphics[width=1\textwidth]{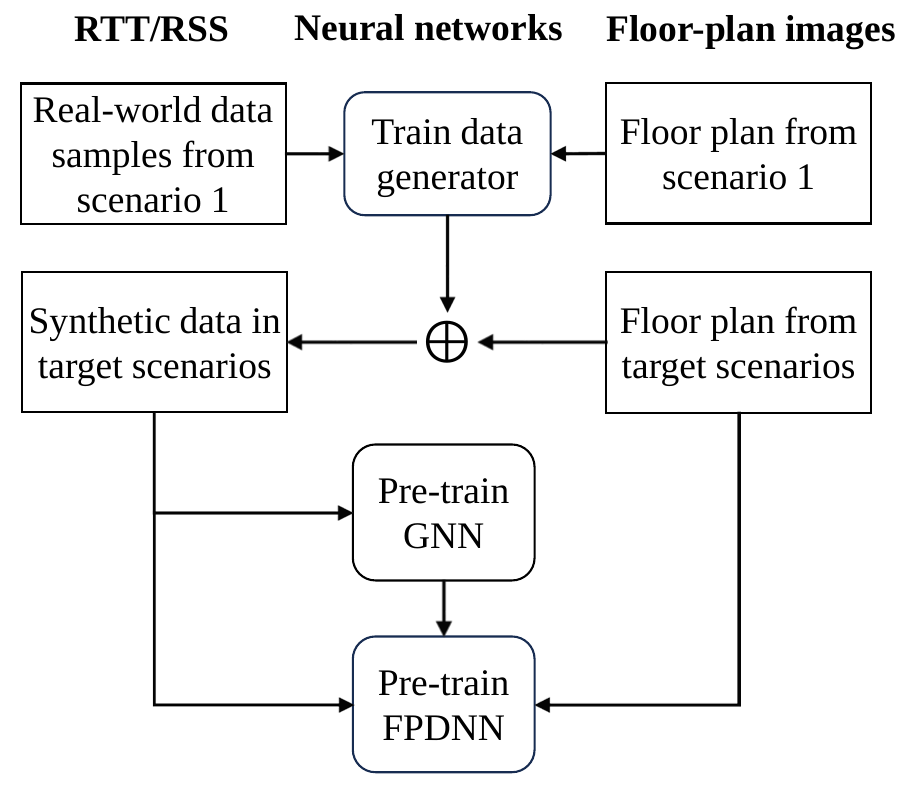}
	\end{minipage}
	%\vspace{-0.1cm}
	\caption{Flow chart of the pre-training stage.}
	\label{fig:flow_chart}
	\vspace{-0.3cm}
\end{figure}

The flow chart of pre-training is illustrated in Fig.~\ref{fig:flow_chart}. Considering that data samples in a real-world environment may not be enough for the training, we establish a virtual environment to generate synthetic data. In different real-world environments, the data samples may follow different distributions. The generator is first trained in the scenario $1$ by the corresponding real-world data and floor-plan image. Then, the generator exploits floor-plan images in different target scenarios to generate synthetic data samples. The GNN and FPDNN are pre-trained offline using synthetic data samples. In this way, we can implement the localization algorithm in unseen environments without real-world data samples.

\begin{figure*}[htbp]
	\centering
	\subfigure[Synthetic RTT.]{
		\begin{minipage}[b]{0.45\textwidth}
			\includegraphics[width=1\textwidth]{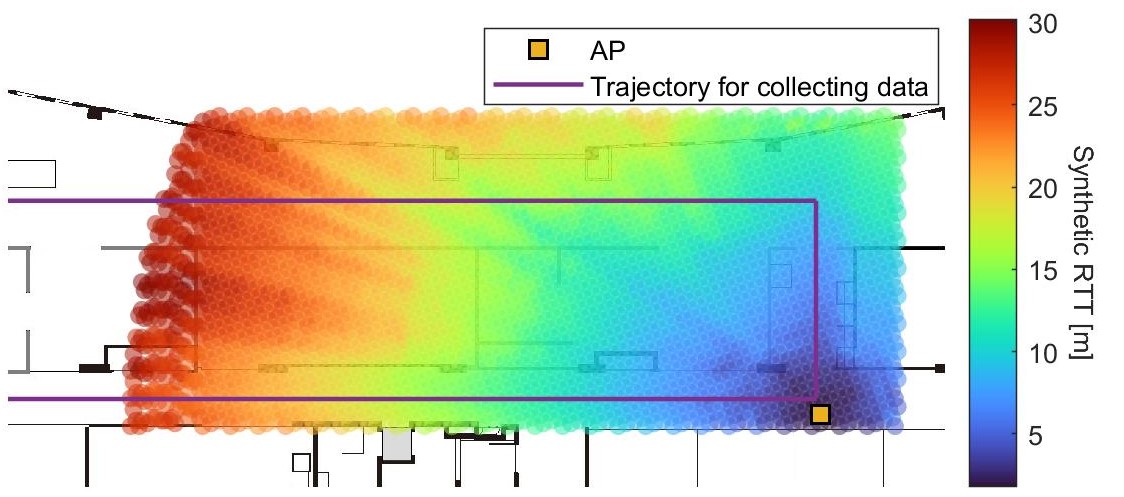} 
		\end{minipage}
	\label{fig:synetic_rtt}
	}
    	\subfigure[Synthetic RSS.]{
    		\begin{minipage}[b]{0.45\textwidth}
   		 	\includegraphics[width=1\textwidth]{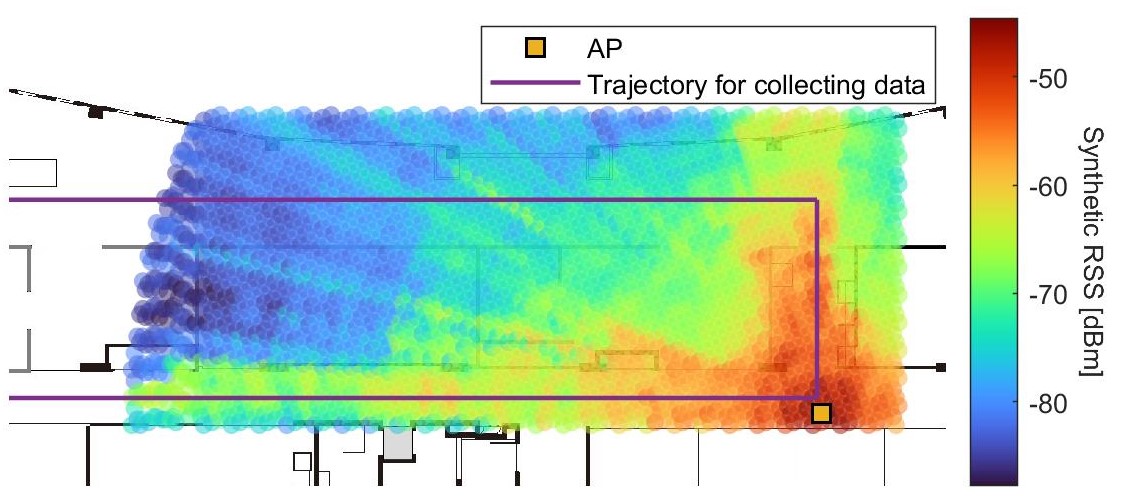}
    		\end{minipage}
		\label{fig:synetic_rss}
    	}
	   
	\caption{Synthetic data generator in the office.}
 \vspace{-0.3cm}
	\label{fig:synethic}
\end{figure*}

\subsection{Synthetic Data Generation in One Virtual Environment}
A data sample of our localization algorithm includes input information and the corresponding label. The input information of the $t$-th sample is composed of node and edge features of a graph, $\mathcal{G}^{(t)}$, and the floor-plan image between the MD and its nearby APs, ${\bf{I}}_{m,k}^{(t)}$. The label is the ground truth location of the MD, ${{\Upsilon}}^{(t)}_m$. 

One way to generate synthetic data samples is to use commercial ray-tracing models \cite{9459462}. The idea is to generate a set of locations of the $m$-th MD and estimate the RTT distance and RSS from each location to its neighbor nodes from ray-tracing models. There are three major issues that forbid the usage of ray-tracing models in our problem: 1) ray-tracing models are very sensitive to the dielectric properties of the reflective surfaces of the obstacles in the radio environment; 2) ray-tracing requires detailed three-dimensional (3D) geometric information; 3) the computing time for generating a sample in ray-tracking models is high, especially when there are a large number of paths between an MD and an AP.

%Ray tracing requires detailed 3-dimensional (3D) geometric information and dielectric properties of the reflective surfaces of the obstacles in the radio environment. While our proposed model uses satellite images and ML to extract the geometric information, which can be deployed in the areas without 3D geometric information. Moreover, the complexity of ray tracing in a 3D environment depends on various factors, such as the number of objects in the scene, the complexity of the geometry, and the number of rays being traced, which is normally high [13]. In contrast, the complexity of the proposed model is fixed since the scale of input features, and the neural network are fixed. In this case, our proposed model can be fast applied in large area.

Alternatively, we use a deep learning model to generate the RTT distance and RSS for a set of locations of the MD. Given the locations of the $m$-th MD and the $k$-th AP, we can obtain the ground-truth distance $d_{m,k}^{(t)}$ and the floor-plan image ${\bf{I}}_{m,k}^{(t)}$ between them. As shown in Fig.~\ref{fig:synetic_data_generator}, the deep learning model is composed of two FFNNs and a DeepVIT and can be expressed as
\begin{equation}
{\relax[\Tilde{\zeta}_{m,k}^{(t)},\Tilde{\gamma}^{(t)}_{m,k}] }= \phi_{\rm D}(d_{m,k}^{(t)},{\bf{I}}_{m,k}^{(t)}; \bm{\theta}_{\rm D}), 
\end{equation}
where $\phi_{\rm D}$ is a DNN in Fig.~\ref{fig:synetic_data_generator}, and the corresponding training parameters are $\bm{\theta}_{\rm D}$. The output $\Tilde{\zeta}_{m,k}$ and $\Tilde{\gamma}_{m,k}$ are the synthetic RTT distance and RSS, respectively.

The generator is trained in a supervised manner with the following loss function
\begin{equation}\label{eq:gradient_rttrss}
\begin{aligned}
  \mathcal{L} ({\bm{\theta}_{\rm D}})=\mathbb{E}((\tilde{\zeta}^{(t)}_{m,k}-{\zeta}^{(t)}_{m,k})^2+(\tilde{\gamma}^{(t)}_{m,k}-{\gamma}^{(t)}_{m,k})^2),
\end{aligned}
\end{equation}
where ${\zeta}^{(t)}_{m,k}$ and ${\gamma}^{(t)}_{m,k}$ are the RTT distance and RSS of the same location in the real-world scenario. For example, to generate synthetic data in the office scenario in Fig.~\ref{fig:synethic}, we collected real-world data samples on the purple trajectory and then used them as labeled samples. After training, we used the synthetic data generator to estimate the RTT distance and RSS at any location in the office.

\subsection{Synthetic Data Generation in Different Environments}
In other environments, such as the shopping mall and laboratory, the corresponding floor-plan images are used to generate synthetic data samples. There is no need to fine-tune the synthetic data generator in new scenarios. Note that the motivation for using synthetic data is to improve the diversity of the training samples for our localization algorithm. The distribution of synthetic data samples in an unseen scenario does not need to be the same as the real-world data samples. The difference between them increases the diversity of data samples and may help improve the generalization ability of our localization algorithm.

\section{WiFi Platform and Data Sets}
In this section, we introduce the WiFi Platform and data sets in each scenario. %Then, the data sets, which include the calibration process, the data set illustration, and the measurement errors of the data set, are introduced. 
\begin{figure*}[htbp]
	%\vspace{-0.1cm}
	\centering
	\begin{minipage}[t]{0.8\textwidth}
		\includegraphics[width=1\textwidth]{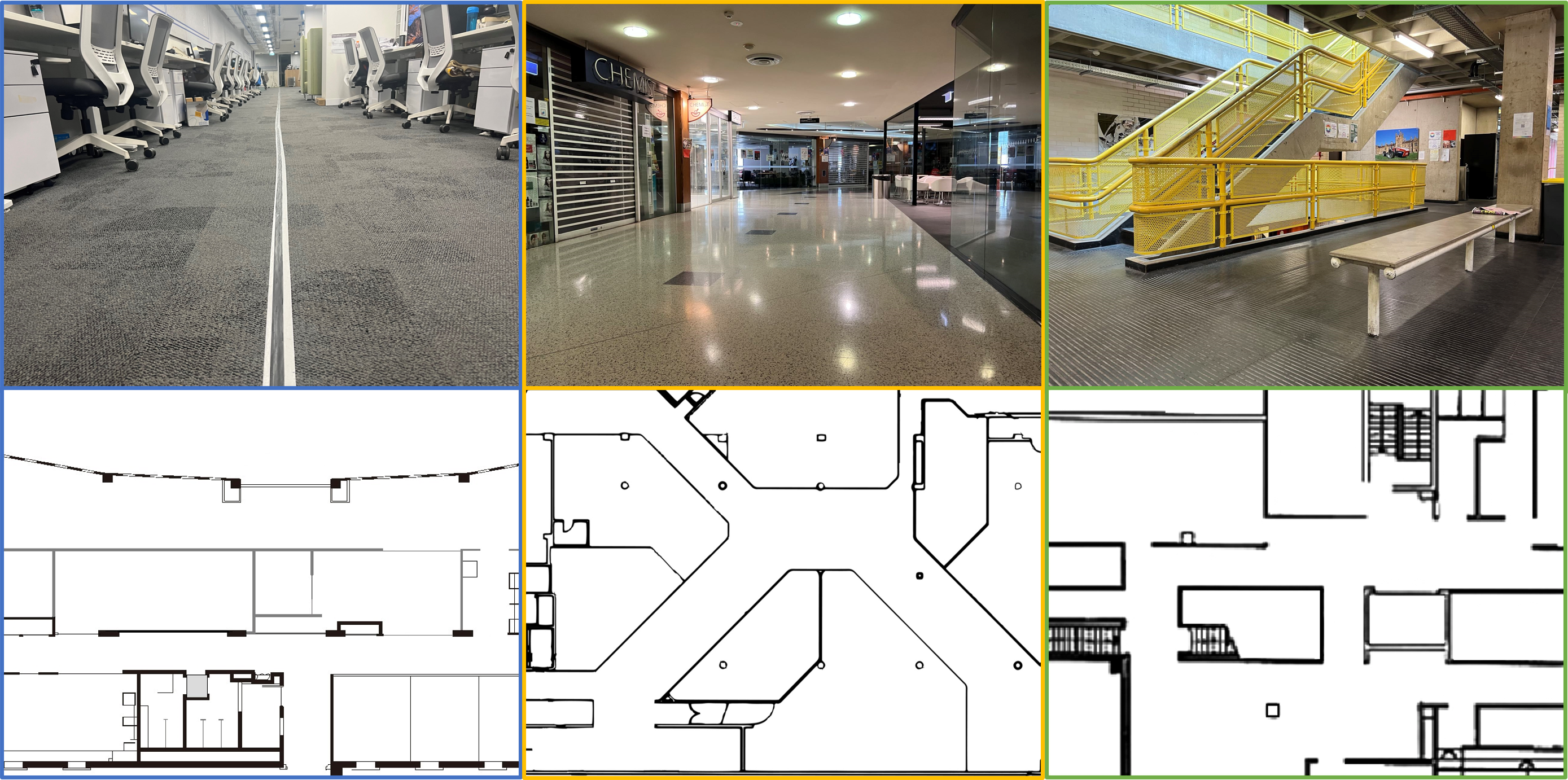}
	\end{minipage}
	%\vspace{-0.1cm}
	\caption{Different environments and corresponding floor plans. Left: Office. Middle: Shopping mall. Right: Laboratory. The furniture is not included in the floor plans, and our evaluation results are obtained in real-world environments with furniture.}
	\label{fig:exp2}
	
\end{figure*}

\subsection{WiFi Platform}
\begin{figure*}[htp]
	%\vspace{-0.1cm}
	\centering
	\begin{minipage}[t]{0.95\textwidth}
		\includegraphics[width=1\textwidth]{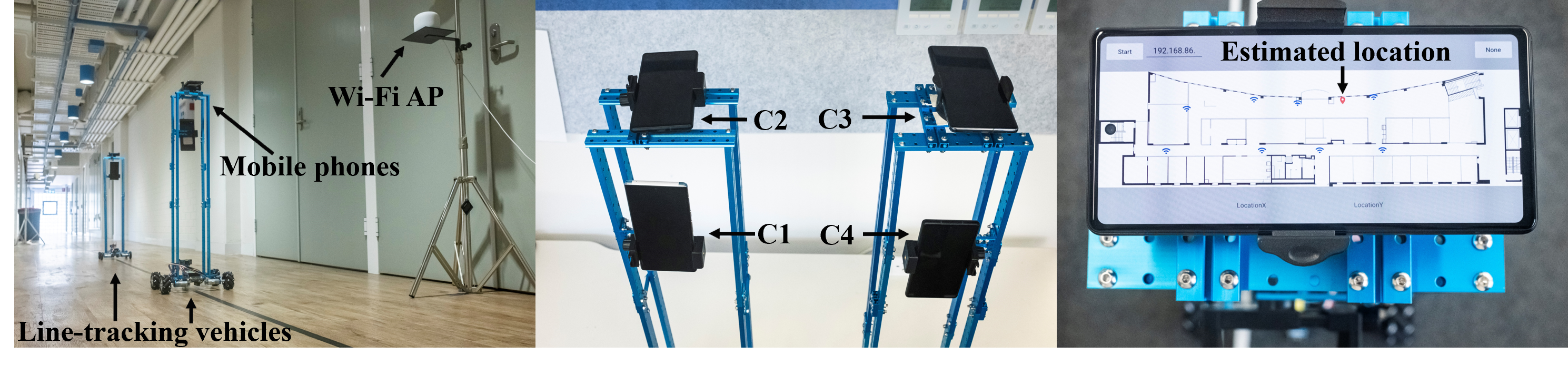}
	\end{minipage}
	%\vspace{-0.1cm}
	\caption{Experiment setup. Left: Wi-Fi AP and line-tracking vehicles. Middle: Mobile devices. Right: estimated location of the MD on the map.}
	\label{fig:exp}
	
\end{figure*}

The WiFi platform consists of eight APs, two line-tracking vehicles, four mobile phones, and a laptop. The details of these devices are given as follows:
\begin{itemize}
\item \textit{Access points}: As shown in Fig.~\ref{fig:exp}, Google Nest WiFi APs, which support both $2.4$~GHz and $5$~GHz with IEEE 802.11 mc standard, are used in our platform. We note that RTT and RSS are obtained from reference signals with carrier frequency at $5$~GHz. In each of the three scenarios, the locations of all eight APs are fixed, and the heights of the tripods are $1.5$~m.  

\item \textit{Line-tracking vehicles}: Makeblock Ultimate 2.0 programmable robot kits are used to build line-tracking vehicles. The line-tracking vehicles are programmed to follow the black line at a constant speed. Each vehicle is equipped with two mobile phones. The horizontal mobile phones (C2 and C3 in Fig.~\ref{fig:exp}) measure the RTT and RSS from nearby APs. The vertical mobile phones (C1 and C4 in Fig.~\ref{fig:exp}) measure the RTT and RSS between each other.

\item \textit{Mobile phones}: The four mobile phones are either Google Pixel 6 Pro or Google Pixel 6a with Android 13. To measure the RTT and RSS between C1 and C4, we use WiFi NanScan, developed by Google. To measure the RTT and RSS from C2 or C3 to all the APs simultaneously, we developed an application ourselves because the available applications can only measure RTT and RSS from one AP to the mobile phone at a time. The sampling interval between the mobile phones and the APs is set to $200$~ms, and the sampling interval between C1 and C4 is set to $1$~s.
\item \textit{Server}: A desktop is used as a server. The detailed specifications for the server are Windows 11 Pro, Intel(R) Core(TM) i9-12900KF central processing unit (CPU) @ 3.20GHz, NVIDIA GeForce RTX 3090, 64GB Memory, and 1TB SSD.
\end{itemize}

\subsection{Typical Scenarios}\label{sec:scenarios}
We collect our data sets in the following three scenarios on the campus of the University of Sydney. All these three scenarios mixed with both LoS and nLoS paths.
\begin{itemize}
\item \textit{Office}: The data samples are collected in the office of the Center of IoT and Telecommunications. This area consists of one long office corridor and an office room, i.e., a $60 \times 20 \;\text{m}^2$ rectangular area.

\item \textit{Laboratory}: The data samples are collected in the Mechanical Engineering Laboratory. The lab has multiple cement posts, iron guardrails, and an elevator, covering a $25 \times 9\;\text{m}^2$ rectangular area.

\item \textit{Shopping mall}: The data samples are collected in the shopping mall in the Wentworth building at the university campus. The experiment is carried out in the mall hallway outside the shops, i.e., a bank, a chemist, and a coffee shop.

\end{itemize}

All the scenarios and their corresponding floor-plan images are provided in Fig.~\ref{fig:exp2}.

\subsection{Data Sets}
The data set\footnote{Available at: https://github.com/haiyaoyu/RTT-Indoor-localization-data-set} of each scenario consists of RTT distance, RSS, and floor-plan image. %In this section, we illustrate our data sets for readers to use.

\subsubsection{Data calibration and processing}\label{data set:data calibration}

As introduced in \cite{ibrahim2018verification}, there are offset components between the RTT distance and the ground truth distance, which lead to inaccurate RTT measurement. These components are static and can be removed from the measured RTT distance. Specifically, we compare the true distance and the measured RTT distance between all the APs and the MD. The results show that the offset components lead to a nearly constant value between the real distance and the measured distance. Then, the average offset between the MD and each AP is removed from the raw RTT distance.

In our data sets, we provide the raw data as well as the data sets after calibration and processing. We note that since the heights of APs and MD are fixed, we convert the RTT distance from in 3D space into a two-dimensional (2D) plane, which can be easily used in 2D localization algorithms. We also preserve the original RTT distance in the 3D space in our data sets.

%To remove the offset components, we measure the real distance and the RTT between all the APs and the MD. \blue{The results show that the offset components lead to a nearly constant gap between measured distance and real distance. Then, the average gap between the MD and each AP is removed from the raw data of RTT..}

%In addition, the heights of the APs and the mobile phones are different. We convert the RTT in a three-dimensional space into the RTT in a two-dimensional plane. After the processing of the RTT samples, it is possible to achieve higher positioning accuracy with low-complexity algorithms.

%In order to access the accurate RTT measurements, the calibration process is introduced. Firstly, the offset, which influenced by different type of hardware, as m, needs to be eliminated. In this case, since there are $2$ different Wi-Fi APs and $3$ different type of UEs, the offset between the real distance and the measured RTT distance in LoS scenario is measured among different type of APs and UEs, respectively. Then, the offset is subtracted from RTT measurements between the corresponding type of UE and AP. 

%Secondly, due to the height difference between APs and UEs, we change the 3 dimension RTT distance to 2 dimension RTT distance for some algorithm, i.e., trilateration method.

\begin{table*}[ht]
\caption{Illustration of Data Samples}
\label{tab:example}
\centering
\resizebox{155mm}{!}{%
\begin{tabular}{|cccccccccc|}
\hline
\multicolumn{10}{|c|}{\textbf{Data samples between a mobile phone and all APs}} \\ \hline
\multicolumn{1}{|c|}{\begin{tabular}[c]{@{}c@{}}Time stamp\\ (ms)\end{tabular}} &
  \multicolumn{1}{c|}{\begin{tabular}[c]{@{}c@{}}AP~$1$ RTT\\ (m)\end{tabular}} &
  \multicolumn{1}{c|}{AP~$1$ RTT std} &
  \multicolumn{1}{c|}{\begin{tabular}[c]{@{}c@{}}AP~$1$ RSS\\ (dBm)\end{tabular}} &
  \multicolumn{1}{c|}{...} &
  \multicolumn{1}{c|}{\begin{tabular}[c]{@{}c@{}}AP~$N$ RTT\\ (m)\end{tabular}} &
  \multicolumn{1}{c|}{AP~$N$ RTT std} &
  \multicolumn{1}{c|}{\begin{tabular}[c]{@{}c@{}}AP~$N$ RSS\\ (dBm)\end{tabular}} &
  \multicolumn{1}{c|}{...} &
  \begin{tabular}[c]{@{}c@{}}Ground Truth\\ $(x_m,y_m)^{(t)}$~(m)\end{tabular} \\ \hline
\multicolumn{1}{|c|}{0} &
  \multicolumn{1}{c|}{1.17} &
  \multicolumn{1}{c|}{0.056} &
  \multicolumn{1}{c|}{-45} &
  \multicolumn{1}{c|}{...} &
  \multicolumn{1}{c|}{14.75} &
  \multicolumn{1}{c|}{0.11} &
  \multicolumn{1}{c|}{-69} &
  \multicolumn{1}{c|}{...} &
  {[}31,13.75{]} \\ \hline
\multicolumn{1}{|c|}{267.10} &
  \multicolumn{1}{c|}{0.58} &
  \multicolumn{1}{c|}{0.055} &
  \multicolumn{1}{c|}{-46} &
  \multicolumn{1}{c|}{...} &
  \multicolumn{1}{c|}{14.27} &
  \multicolumn{1}{c|}{0.055} &
  \multicolumn{1}{c|}{-68} &
  \multicolumn{1}{c|}{...} &
  {[}31.03,13.75{]} \\ \hline
\multicolumn{1}{|c|}{...} &
  \multicolumn{1}{c|}{...} &
  \multicolumn{1}{c|}{...} &
  \multicolumn{1}{c|}{...} &
  \multicolumn{1}{c|}{...} &
  \multicolumn{1}{c|}{...} &
  \multicolumn{1}{c|}{...} &
  \multicolumn{1}{c|}{...} &
  \multicolumn{1}{c|}{...} &
  ... \\ \hline

\end{tabular}%
}
\end{table*}

\subsubsection{Illustration of data set}
As shown in Section~\ref{sec:scenarios}, we collected our data samples in three scenarios. For each scenario, we created four files: 1) the floor plan of the scenario, 2) measured information from each mobile phone to all the APs, 3) measured information between C1 and C4, and 4) the locations and indices of the APs.

We take the “office” as an example to illustrate our data sets. The locations of all the APs, the trajectory of the vehicles, and the layout of the office can be found in Fig.~\ref{fig:data_train_generator}. As shown in Table~\ref{tab:example}, a sample between the MD and all APs consists of a time stamp, RTT distance, RSS, and the ground truth location of the MD.

\subsection{Measurement Errors}

\subsubsection{RSS measurement errors}
From the theoretical path-loss models of wireless channels, we can estimate the distance between two devices based on the RSS. The RSS could be inaccurate due to stochastic interference and noise. In addition, since the communication environment is dynamic,  walls and obstacles may cause severe signal attenuation and contribute to additional estimation errors \cite{8924707}. As a result, the communication distance estimated from the RSS is inaccurate.

\subsubsection{RTT measurement errors}\label{RTTerroranalysis}
Based on the assumption that RTT equals the propagation time, it is possible to estimate the distance between two devices based on the RTT. However, the assumption does not hold in practice. Specifically, the unstable and low-precision oscillators may cause clock measurement noise. In addition, due to the Media Access Control (MAC) processing delay, the multi-path effect, refractive indices of different propagation mediums, and other offset components caused by antennas and chips, the measured RTT is longer than the propagation time \cite{8924707}. It is worth noting that the offset components are static, and it is possible to remove this part from the raw data. %\cite{ibrahim2018verification}. 

\section{Experimental Results}
In this section, we present extensive experimental results to verify the accuracy of our proposed indoor localization system, and the performance of our proposed algorithm is further compared with some existing baseline methods in terms of localization error. First, the hyper-parameters of various blocks in the proposed model and the existing baselines are summarized. Then, we evaluate the localization accuracy of the proposed models in the office scenario in Section~\ref{samearea}. Next, we extend to deploy the proposed framework in the shopping mall and laboratory in Section~\ref{different_location}. Finally, we evaluate our proposed algorithm in the multi-user scenario in Section~\ref{multi-user-condition}.

\subsection{Hyper-Parameters}\label{numerical}
\subsubsection{FPDNN}

The hyper-parameters of each block, e.g., FFNN, DeepVIT, of our proposed FPDNN model are summarized in Table~\ref{tab:parameter}. We use MSE as the loss function, \textit{Adam} as the optimizer, with a learning rate of $1\times 10^{-3}$, and a weight decay of $1\times 10^{-5}$ \cite{kingma2014adam}. During the training of the FPDNN, the epoch is set to $50$, and the batch size is set to $32$. 

\subsubsection{GNN}
The hyper-parameters of different steps in GNN, e.g., message-passing and aggregation, are summarized in Table~\ref{tab:parameter}. We note that the loss function, the training epochs, and the batch size are the same as FPDNN. We use the Adam optimizer with a learning rate $2\times 10^{-4}$.

\subsubsection{DNN in synthetic data generator}
We note that the DeepVIT block and the FFNN3 in the synthetic data generation DNN have the same structure as the DeepVIT and FFNN1 in FPDNN, respectively. Thus, we omit the details of them in Table~\ref{tab:parameter}. In addition, the loss function, the optimizer, the number of epochs, and the batch size are the same as FPDNN.

% \usepackage{multirow}
% \usepackage{graphicx}
\begin{table*}[]
\centering
\caption{The Hyper-parameters for The GNN Model, FPDNN Model, and Data Generator.}
\label{tab:parameter}
\resizebox{175mm}{!}{%
\begin{tabular}{|ccccc|}
\hline
\multicolumn{5}{|c|}{\textbf{FPDNN}} \\ \hline
\multicolumn{1}{|c|}{\textbf{Block}} &
  \multicolumn{1}{c|}{\textbf{Layer}} &
  \multicolumn{1}{c|}{\textbf{Input Shape}} &
  \multicolumn{1}{c|}{\textbf{Output Shape}} &
  \textbf{Remark} \\ \hline
\multicolumn{1}{|c|}{\multirow{3}{*}{FFNN1}} &
  \multicolumn{1}{c|}{FC1} &
  \multicolumn{1}{c|}{2} &
  \multicolumn{1}{c|}{64} &
  first fully-connected layer with Relu \\ \cline{2-5} 
\multicolumn{1}{|c|}{} &
  \multicolumn{1}{c|}{Norm} &
  \multicolumn{1}{c|}{64} &
  \multicolumn{1}{c|}{64} &
  layer normalization \\ \cline{2-5} 
\multicolumn{1}{|c|}{} &
  \multicolumn{1}{c|}{FC2} &
  \multicolumn{1}{c|}{64} &
  \multicolumn{1}{c|}{32} &
  second fully-connected layer with Relu and 0.2 dropout rate \\ \hline
\multicolumn{1}{|c|}{\multirow{3}{*}{Image Pre-processing}} &
  \multicolumn{1}{c|}{Image Splitting} &
  \multicolumn{1}{c|}{$W^{(t)}\times 256\times1$} &
  \multicolumn{1}{c|}{$N_s^{(t)}\times 32\times 32$} &
  $H = 256, C = 1, P = 32$ \\ \cline{2-5} 
\multicolumn{1}{|c|}{} &
  \multicolumn{1}{c|}{Flattening} &
  \multicolumn{1}{c|}{$N_s^{(t)}\times 32\times 32$} &
  \multicolumn{1}{c|}{$N_s^{(t)}\times 512$} &
  $D' = 1024, D = 512$ \\ \cline{2-5} 
\multicolumn{1}{|c|}{} &
  \multicolumn{1}{c|}{Position Embedding} &
  \multicolumn{1}{c|}{$N_s^{(t)}\times 512$} &
  \multicolumn{1}{c|}{$(N_s^{(t)}+1)\times 512$} &
  add extra learnable embedding \\ \hline
\multicolumn{1}{|c|}{\multirow{6}{*}{\begin{tabular}[c]{@{}c@{}}DeepVIT\\ Transformer Encoder\end{tabular}}} &
  \multicolumn{1}{c|}{Norm} &
  \multicolumn{1}{c|}{$(N_s^{(t)}+1)\times 512$} &
  \multicolumn{1}{c|}{$(N_s^{(t)}+1)\times 512$} &
  layer normalization \\ \cline{2-5} 
\multicolumn{1}{|c|}{} &
  \multicolumn{1}{c|}{Re-attention} &
  \multicolumn{1}{c|}{$(N_s^{(t)}+1)\times 512$} &
  \multicolumn{1}{c|}{$2\times (N_s^{(t)})\times 64$} &
  number of attention heads: $2$ \\ \cline{2-5} 
\multicolumn{1}{|c|}{} &
  \multicolumn{1}{c|}{FC1} &
  \multicolumn{1}{c|}{$2\times (N_s^{(t)})\times 64$} &
  \multicolumn{1}{c|}{$N_s^{(t)}\times 512$} &
  first fully-connected layer with Relu \\ \cline{2-5} 
\multicolumn{1}{|c|}{} &
  \multicolumn{1}{c|}{Norm} &
  \multicolumn{1}{c|}{$N_s^{(t)}\times 512$} &
  \multicolumn{1}{c|}{$N_s^{(t)}\times 512$} &
  layer normalization \\ \cline{2-5} 
\multicolumn{1}{|c|}{} &
  \multicolumn{1}{c|}{FC2} &
  \multicolumn{1}{c|}{$N_s^{(t)}\times 512$} &
  \multicolumn{1}{c|}{$N_s^{(t)}\times 512$} &
  a fully-connected layer with Gelu and 0.2 dropout rate \\ \cline{2-5} 
\multicolumn{1}{|c|}{} &
  \multicolumn{1}{c|}{mean} &
  \multicolumn{1}{c|}{$N_s^{(t)}\times 512$} &
  \multicolumn{1}{c|}{$1\times 512$} &
  - \\ \hline
\multicolumn{1}{|c|}{\multirow{2}{*}{\begin{tabular}[c]{@{}c@{}}DeepVIT\\ FFNN Head\end{tabular}}} &
  \multicolumn{1}{c|}{Norm} &
  \multicolumn{1}{c|}{$1\times 512$} &
  \multicolumn{1}{c|}{$1\times 512$} &
  layer normalization \\ \cline{2-5} 
\multicolumn{1}{|c|}{} &
  \multicolumn{1}{c|}{FC1} &
  \multicolumn{1}{c|}{$1\times 512$} &
  \multicolumn{1}{c|}{$1\times 32$} &
  fully-connected layer with Relu \\ \hline
\multicolumn{1}{|c|}{\multirow{3}{*}{FFNN2}} &
  \multicolumn{1}{c|}{FC1} &
  \multicolumn{1}{c|}{$1\times 64$} &
  \multicolumn{1}{c|}{$1\times 32$} &
  first fully-connected layer with Relu \\ \cline{2-5} 
\multicolumn{1}{|c|}{} &
  \multicolumn{1}{c|}{Norm} &
  \multicolumn{1}{c|}{$1\times 32$} &
  \multicolumn{1}{c|}{$1\times 16$} &
  layer normalization \\ \cline{2-5} 
\multicolumn{1}{|c|}{} &
  \multicolumn{1}{c|}{FC2} &
  \multicolumn{1}{c|}{$1\times 16$} &
  \multicolumn{1}{c|}{1} &
  second fully-connected layer with Relu \\ \hline
\multicolumn{5}{|c|}{\textbf{GNN}} \\ \hline
\multicolumn{1}{|c|}{\multirow{2}{*}{\begin{tabular}[c]{@{}c@{}}Message-passing\\ $\phi(\cdot;\theta)$\end{tabular}}} &
  \multicolumn{1}{c|}{FC1} &
  \multicolumn{1}{c|}{9} &
  \multicolumn{1}{c|}{8} &
  fully-connected layer with Relu \\ \cline{2-5} 
\multicolumn{1}{|c|}{} &
  \multicolumn{1}{c|}{FC2} &
  \multicolumn{1}{c|}{8} &
  \multicolumn{1}{c|}{2} &
  second fully-connected layer  \\ \hline
\multicolumn{1}{|c|}{Aggregation} &
  \multicolumn{1}{c|}{mean} &
  \multicolumn{1}{c|}{2} &
  \multicolumn{1}{c|}{2} &
  - \\ \hline
\multicolumn{1}{|c|}{\multirow{2}{*}{\begin{tabular}[c]{@{}c@{}}Update\\ $\rm{U}(\cdot;\varphi)$\end{tabular}}} &
  \multicolumn{1}{c|}{FC1} &
  \multicolumn{1}{c|}{7} &
  \multicolumn{1}{c|}{8} &
  first fully-connected layer with Relu \\ \cline{2-5} 
\multicolumn{1}{|c|}{} &
  \multicolumn{1}{c|}{FC2} &
  \multicolumn{1}{c|}{8} &
  \multicolumn{1}{c|}{2} &
  second fully-connected layer with Relu \\ \hline
\multicolumn{5}{|c|}{\textbf{DNN in Synthetic Data Generation}} \\ \hline
\multicolumn{1}{|c|}{\multirow{3}{*}{FFNN4}} &
  \multicolumn{1}{c|}{FC1} &
  \multicolumn{1}{c|}{64} &
  \multicolumn{1}{c|}{32} &
  first fully-connected layer with Relu \\ \cline{2-5} 
\multicolumn{1}{|c|}{} &
  \multicolumn{1}{c|}{Norm} &
  \multicolumn{1}{c|}{32} &
  \multicolumn{1}{c|}{32} &
  layer normalization \\ \cline{2-5} 
\multicolumn{1}{|c|}{} &
  \multicolumn{1}{c|}{FC2} &
  \multicolumn{1}{c|}{32} &
  \multicolumn{1}{c|}{2} &
  second fully-connected layer with Relu and 0.2 dropout rate \\ \hline
\end{tabular}%
}
\end{table*}

\subsection{Other Existing Baselines}\label{baseline}

We stimulate some existing methods as baselines to compare with our proposed indoor localization system. The baseline methods are:

% \subsubsection{Trilateration}

% This conventional localization system uses the trilateration algorithm \cite{li2017novel} to estimate the location of the MD based on the raw RTT distance from the three nearest APs. 

\subsubsection{Least square}
We use the least square algorithm \cite{4212819} to estimate the location of the MD  based on the raw RTT distance from all collected APs, as introduced in Section~\ref{LS_algorithm}. 

\subsubsection{RTT-RSS model}
In \cite{8924707}, the authors designed a hybrid RTT-RSS model for localization, which integrated RTT, RSS, and Kalman filter to enhance the scalability and robustness of the positioning system. 

\subsubsection{Bayesian grid update method}
We reproduce the algorithm from~\cite{horn2022indoor}, which utilizes the Bayesian data analysis to enhance the positioning accuracy.

\subsubsection{CbT \& WCCG}
We also reproduce the algorithm from \cite{9151400}, which utilizes the clustering-based trilateration (CbT) and the weighted concentric circle generation (WCCG) to estimate the location of the MD.

% This conventional localization system uses the trilateration algorithm \cite{li2017novel} to estimate the location of the MD based on the raw RTT distance from the three nearest APs. 
% \subsubsection{Models from other literature}
% We also reproduce two recently published algorithms based on our data set. The first one is from \cite{8924707}, where the authors designed a hybrid RTT-RSS model for localization. The second one is from \cite{horn2022indoor} based on the Bayesian grid update.

\subsection{Performance Metrics}
The localization error of the $m$-th MD at the $t$-th time slot is defined as
\begin{equation}
\begin{aligned}
\epsilon_m^{(t)}=\sqrt{(\hat{x}_m^{(t)}-x^{(t)}_{m})^2+(\hat{y}_m^{(t)}-y^{(t)}_{m})^2}.
\end{aligned}
\end{equation}
 From the cumulative distribution function (CDF) of $\epsilon_m^{(t)}$, we can obtain the mean absolute error (MAE) (with legend ``MAE"), the median of the errors (with legend ``$50$\% CDF"), and the 90-th percentile CDF (with legend ``$90$\% CDF"). In addition, we can evaluate the RMSE of the errors according to the following expression
\begin{equation}
\begin{aligned}
E_m=\sqrt{\frac{1}{N_t}\sum_{t=1}^{N_t} (\epsilon_m^{(t)})^2},
\end{aligned}
\end{equation}
where $N_t$ is the number of testing samples.

\subsection{Training of Synthetic Data Generator in The Office}\label{Training of Synthetic Data Generator in the Office}

\begin{figure}[htp]
	%\vspace{-0.1cm}
	\centering
	\begin{minipage}[t]{0.45\textwidth}
		\includegraphics[width=1\textwidth]{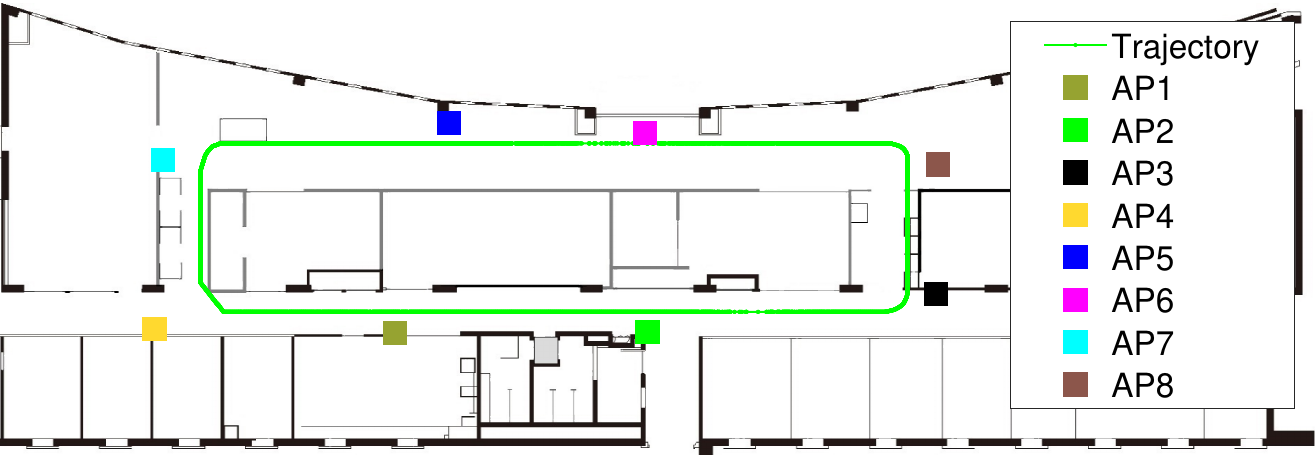}
	\end{minipage}
	%\vspace{-0.1cm}
	\caption{Locations of eight APs and MD, where the MD move along the trajectory in the real-world scenario.}
	\label{fig:data_train_generator}
	\vspace{-0.3cm}
\end{figure}
To train the synthetic data generator, we collect samples in the office. As shown in Fig.~\ref{fig:data_train_generator}, a line-tracking robot moves along the trajectory (with legend ``Trajectory") for five circles. We have $50,000$ data samples in this scenario, and each data sample consists of the RTT and RSS information from an AP to the MD. We use the samples collected in the first four circles to train the data generator. After the training, the generator can output RTT and RSS information at any location in the office, as illustrated in Fig.~\ref{fig:synethic}. %In this case, the GNN and FPDNN can be pre-trained by the synthetic data in the virtual environment.

\subsection{Localization in The Office}\label{samearea}
\subsubsection{Training with real-world data samples}
\begin{figure}[htbp]
	\centering
    	
	\subfigure[Estimated locations with RTT-RSS model and Bayesian grid update.]{
    		\begin{minipage}[b]{0.45\textwidth}
   		 	\includegraphics[width=1\textwidth]{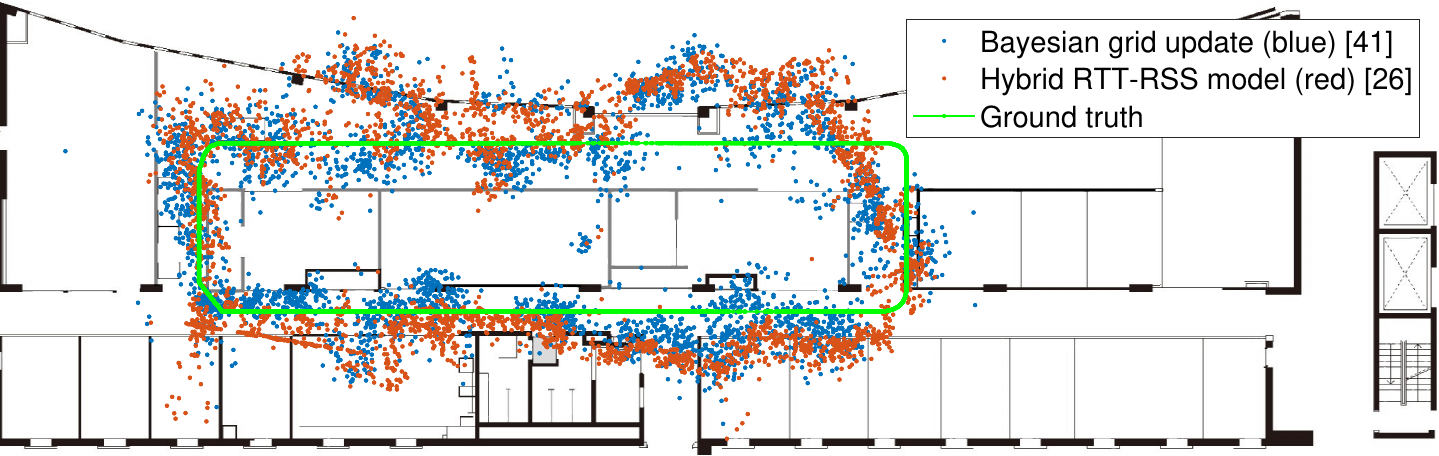}
    		\end{minipage}
		\label{fig:same_scenario_traj_2_baselines_1}
    	}
     \subfigure[Estimated locations with CbT+WCCG and proposed GNN+FPDNN .]{
    		\begin{minipage}[b]{0.45\textwidth}
   		 	\includegraphics[width=1\textwidth]{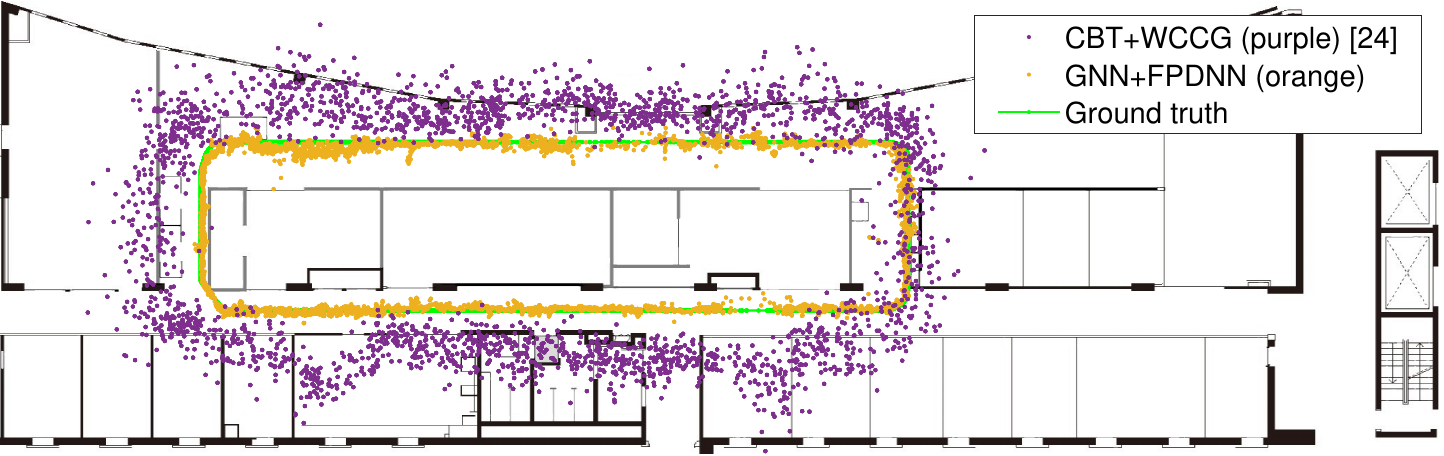}
    		\end{minipage}
		\label{fig:same_scenario_traj_2_baselines_2}
    	}

	\caption{Localization with our method (GNN+FPDNN) and three existing baselines.}
	\label{fig:same_scenario_traj}
 \vspace{-0.3cm}
\end{figure}

We first evaluate the performance of our algorithm when it is trained with real-world data samples in the office environment. Specifically, we use $80$\% of the samples for training and the rest $20$\% of the samples for testing. The estimated locations using the hybrid RTT-RSS model in \cite{8924707}, Bayesian grid update model in \cite{horn2022indoor}, CbT+WCCG method in \cite{9151400}, and our proposed method, are shown in Fig.~\ref{fig:same_scenario_traj}. The results indicate that the estimated locations of our method are well-aligned with the ground-truth trajectory in Fig.~\ref{fig:data_train_generator}, and the localization errors are much lower than the other three baselines.

% trajectory of the MD estimated by the hybrid RTT-RSS model and the Bayesian grid update falls outside the ground truth trajectory and has tremendous variation, with the effect of nLoS paths. The estimated trajectory of the MD for our proposed algorithm observes remarkable results, in which the estimated trajectory almost matches the ground truth trajectory.

% \begin{figure}[htp]
% 	%\vspace{-0.1cm}
% 	\centering
% 	\begin{minipage}[t]{0.5\textwidth}
% 		\includegraphics[width=1\textwidth]{draft/same_scenario_traj_3_baselines.pdf}
% 	\end{minipage}
% 	%\vspace{-0.1cm}
% 	\caption{Localization with our method (GNN+FPDNN) and two existing baselines.}
% 	\label{fig:same_scenario_traj}
% 	\vspace{-0.3cm}
% \end{figure}

 %We only observe a few points falling outside the ground truth trajectory, which could be caused by radio interference from other equipment.

\begin{figure}[htp]
	\vspace{-0.1cm}
	\centering
	\begin{minipage}[t]{0.45\textwidth}
		\includegraphics[width=1\textwidth]{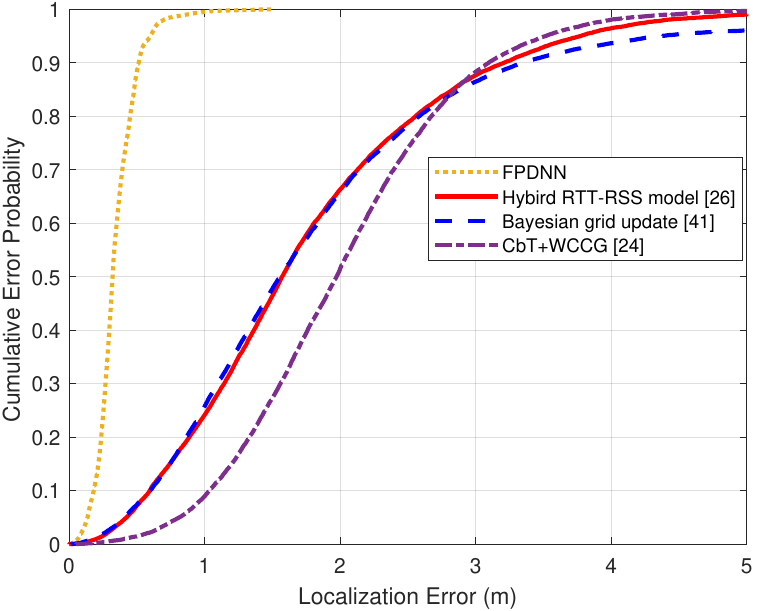}
	\end{minipage}
	\vspace{-0.1cm}
	\caption{CDFs of localization errors, where GNN and FPDNN are trained with real-world samples.}
	\label{fig:8apcdf}
	\vspace{-0.3cm}
\end{figure}

To further illustrate the localization errors of different methods, we provide the CDFs of localization errors in Fig.~\ref{fig:8apcdf}. With our algorithm, the median of localization errors is $0.2$~m, and the localization errors are lower than $0.44$~m with a probability of $90$\%. For the other three baselines, their medians are larger than $1.5$~m, and the $90$-th percentile CDFs are larger than $3$~m.

%As mentioned above, the 

\begin{figure}[htbp]
	\centering
    	
	\subfigure[Pre-localization with GNN.]{
    		\begin{minipage}[b]{0.45\textwidth}
   		 	\includegraphics[width=1\textwidth]{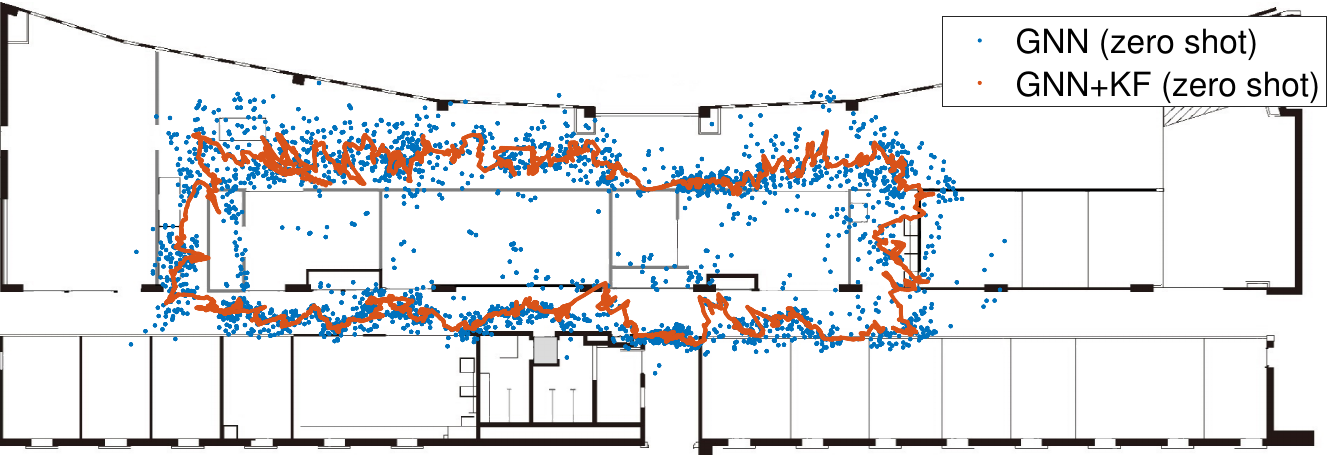}
    		\end{minipage}
		\label{fig:after_GNN}
    	}
     \subfigure[Estimated locations with GNN+FPDNN.]{
    		\begin{minipage}[b]{0.45\textwidth}
   		 	\includegraphics[width=1\textwidth]{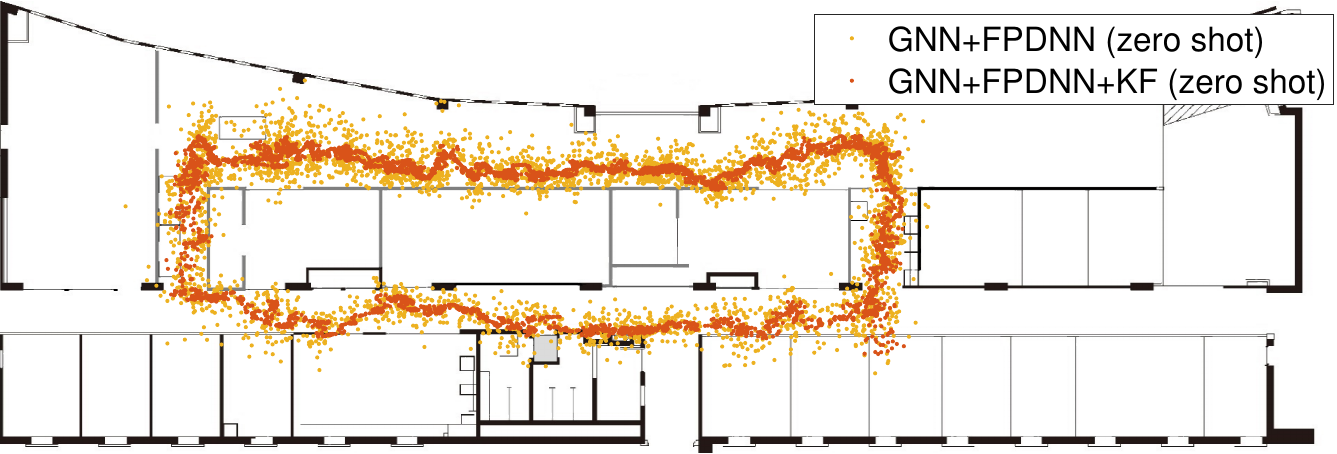}
    		\end{minipage}
		\label{fig:after_FPDNN}
    	}

	\caption{Locations of APs, ground truth, and estimated locations.}
	\label{fig:different_pro_scenario}
 \vspace{-0.3cm}
\end{figure}

\subsubsection{Zero-shot learning without real-world data samples}\label{different_scenario}
\begin{figure}[hthp]
	\vspace{-0.1cm}
	\centering
	\begin{minipage}[t]{0.45\textwidth}		\includegraphics[width=1\textwidth]{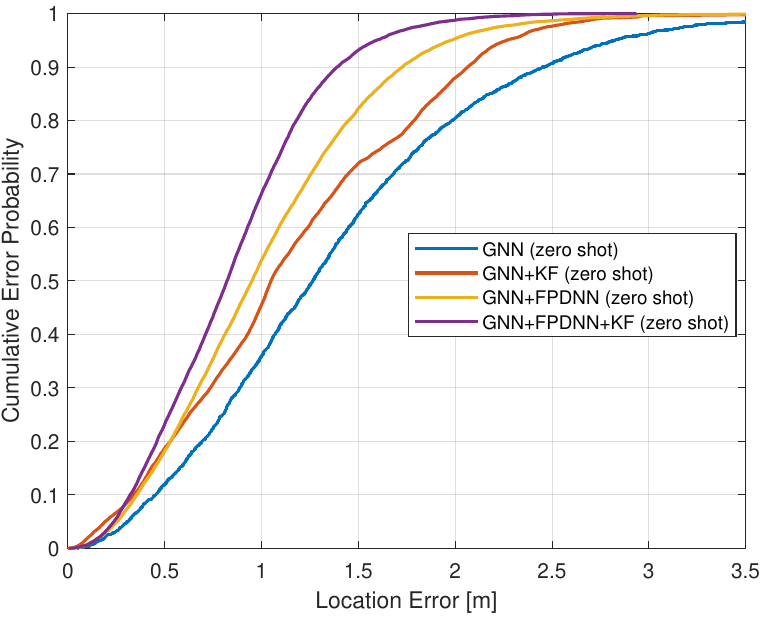}
	\end{minipage}
	\vspace{-0.1cm}
	\caption{CDFs of localization errors achieved by zero-shot learning approaches.}
	\label{fig:cdf_dif_sce}
	\vspace{-0.3cm}
\end{figure}

\begin{figure}[tbp]
	\centering
    	\subfigure[Real-world data samples for training/testing of localization algorithm.]{
    		\begin{minipage}[b]{0.45\textwidth}
   		 	\includegraphics[width=1\textwidth]{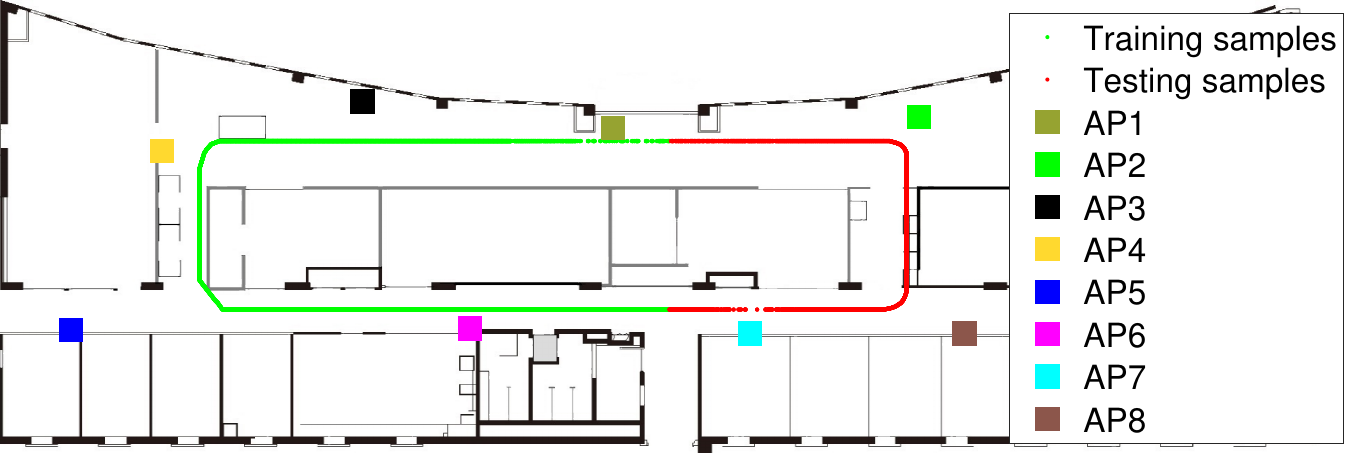}
    		\end{minipage}
		\label{fig:propo_scenario_online}
    	}
	\subfigure[Estimated location with fine-tuning.]{
    		\begin{minipage}[b]{0.45\textwidth}
   		 	\includegraphics[width=1\textwidth]{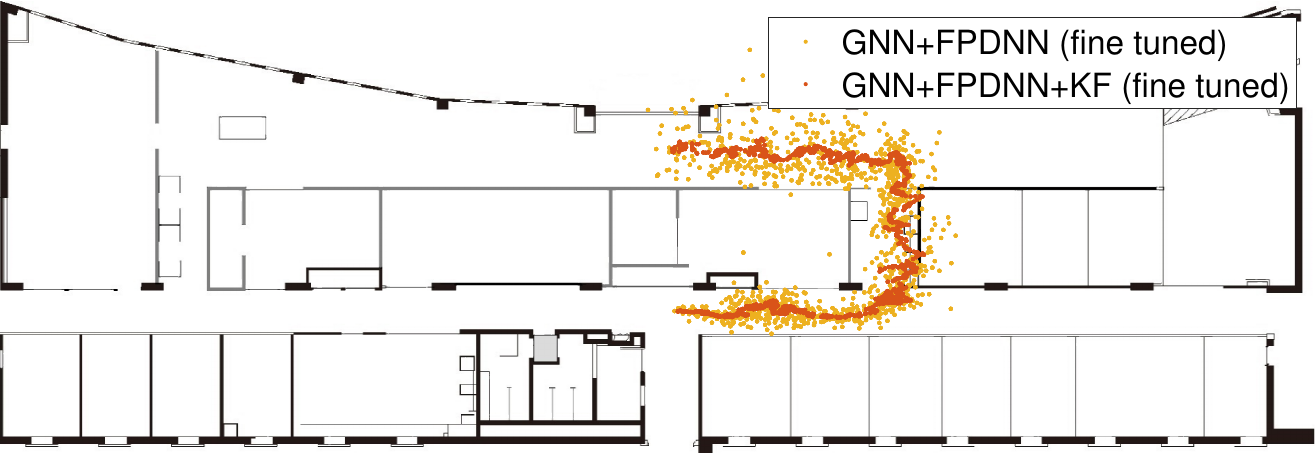}
    		\end{minipage}
		\label{fig:after_FPNN_fine_tuned}
    	}
	\caption{Locations of APs, ground truth, and estimated locations.}	\label{fig:different_pro_scenario_fine_tuned}
 \vspace{-0.3cm}
\end{figure}

In Figs. \ref{fig:different_pro_scenario} and \ref{fig:cdf_dif_sce}, GNN and FPDNN are trained with the synthetic data samples. To validate the effectiveness of our synthetic data generator, we changed the deployment of APs in testing. The locations of APs in Fig. \ref{fig:data_train_generator} are different from the locations of APs in Fig. \ref{fig:propo_scenario_online}.  Since there is no real-world sample in the new scenario, this approach is known as zero-shot learning.

% we consider different trajectories for training and evaluation. Specifically, the data generator is trained by the same data samples and scenario in Section~\ref{Training of Synthetic Data Generator in the Office}. Then, we change the location of all APs to ensure the propagation scenario of each AP is different from the previous one for evaluation, as shown in Fig.~\ref{fig:propo_scenario_online}. We note that the pre-trained GNN and FPDNN are evaluated without any data samples collected in the new environment, which is named "zero-shot".

%trajectory of the MD during the two scenarios is the same; while the locations of APs are different (the same AP present in the same color in Fig.~\ref{fig:data_train_generator} and Fig.~\ref{fig:propo_scenario_online}). %In the second scenario, the trajectory in green color can be used for fine-tuning the system, and the red color is used for testing.

The estimated locations in Fig.~\ref{fig:after_GNN} are obtained with the GNN (either with or without Kalman Filter). With the help of the Kalman Filter, we can obtain better pre-localization accuracy, and the corresponding results serve as one part of the input in the FPDNN. In Fig.~\ref{fig:after_FPDNN}, we show the final results obtained from the FPDNN. The results indicate that by combining the GNN, the FPDNN, and the Kalman filter, the estimated locations are close to the ground truth in Fig.~\ref{fig:data_train_generator}. The RMSE of the estimated location with the GNN and Kalman filter is $1.33$~m, and the location RMSE with the GNN+FPDNN is $1.15$~m. After smoothing by the Kalman filter, the RMSE further decreases to $0.95$~m.

The CDFs of the location errors are illustrated in Fig.~\ref{fig:cdf_dif_sce}. From the CDFs, the medians of localization errors achieved by the four approaches mentioned above are $0.82$~m, $0.95$~m, $1.05$~m, and $1.26$~m, respectively. With a probability of $90$\%, the estimation errors of the four approaches are lower than $1.39$~m, $1.73$~m, $2.06$~m, and $2.46$~m, respectively.
By combining the GNN, the FPDNN, and the Kalman Filter, the RMSE is $0.95$~m, even with no sample in the new scenario. These results indicate that with the help of synthetic data samples, our method outperforms the three existing baselines in unseen scenarios. 

%The experiment results indicate that the FPDNN, which utilizes the floor plan image, can further improve the accuracy. What is more, the Kalman filter, which utilizes temporal information, can further increase the localization accuracy. Finally, with the help of the synthetic data generator, the proposed indoor localization system can achieve sub-meter accuracy when deployed in different propagation scenarios. 

\begin{figure*}[tbp]
 
	\centering
	\subfigure[Training with real-world data samples.]{
		\begin{minipage}[b]{0.4\textwidth}
			\includegraphics[width=1\textwidth]{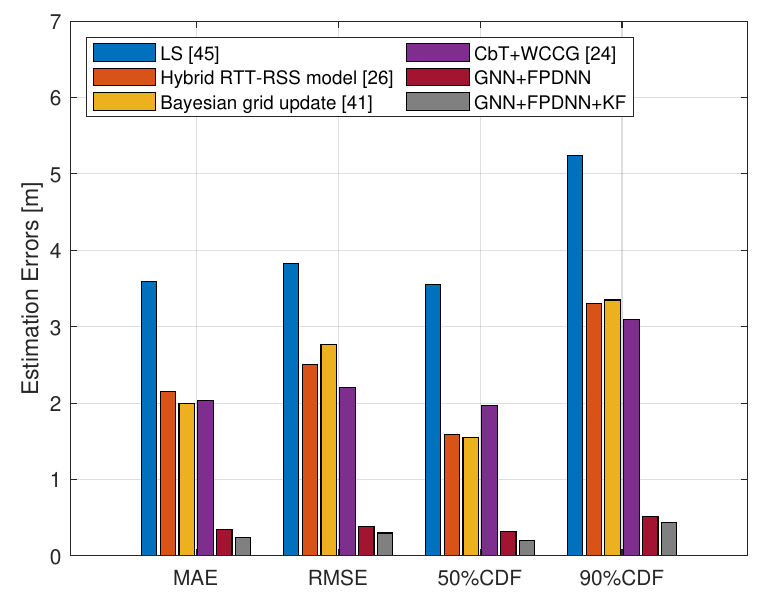} 
		\end{minipage}
		
		\label{fig:same_scenario_bar}
	}
    	\subfigure[Training with synthetic data samples, and fine-tuning with part of real-world data samples.]{
    		\begin{minipage}[b]{0.4\textwidth}
   		 	\includegraphics[width=1\textwidth]{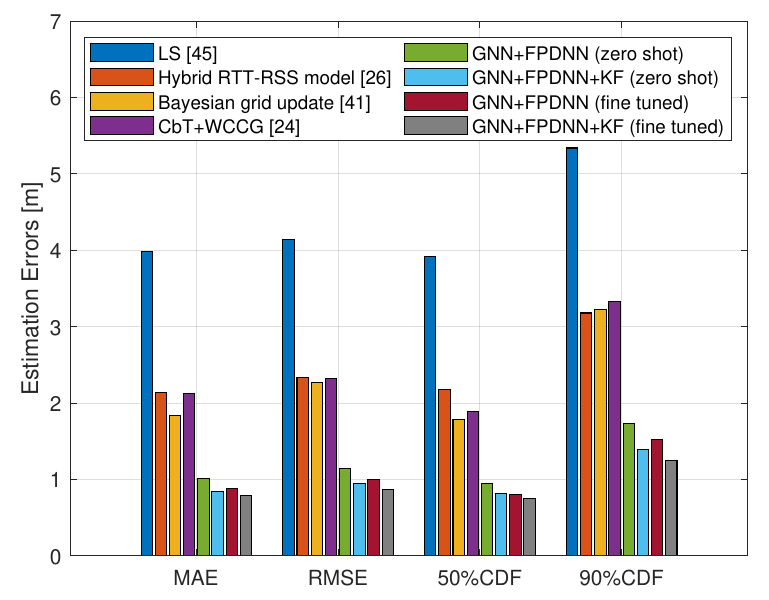}
    		\end{minipage}
		\label{fig:different_scenario_bar}
    	}
     \vspace{-0.1cm}
	\caption{Comparison of different methods.}
	\label{fig:office_experiment}
  \vspace{-0.3cm}
\end{figure*}

\subsubsection{Fine-tuning with part of real-world data samples} 
The localization accuracy can be further improved if parts of the real-world data samples are used to fine-tune the pre-trained GNN and FPDNN. As shown in Fig.~\ref{fig:propo_scenario_online}, the samples used in fine-tuning and testing are in green and red, respectively. The testing results are shown in Fig.~\ref{fig:after_FPNN_fine_tuned}. To better illustrate the gaps among different approaches, we provide the MAE, RMSE, median, and $90$-th percentile CDF in Fig.~\ref{fig:office_experiment}. The results indicate that by fine-tuning the GNN and FPDNN, it is possible to obtain $10$\% $\sim 20$\% performance gain compared with zero-shot learning approaches. For example, the localization RMSE is $1.15$~m for the zero-shot GNN and FPDNN, while the RMSE decreases to $1$~m after fine-tuning. Nevertheless, the localization errors achieved by the zero-shot learning approaches are lower than the baselines.

By comparing Figs. \ref{fig:same_scenario_bar} and \ref{fig:different_scenario_bar}, we can see that if we only have part of the environment information, there is a performance loss. If the GNN and FPDNN are trained with data samples 
obtained in the first four circles and tested with the data samples obtained in the last circle, the RMSE is $0.3$~m (with GNN, FPDNN, and Kalman filter). If we train the GNN and FPDNN with the samples in green and test them with red samples, then the RMSE is $0.87$~m (with GNN, FPDNN, and Kalman filter). This is because the green samples do not provide the environment information in the right part of the office. 

\begin{figure*}[tbp]
	\centering
	\subfigure[Estimation errors in the shopping mall.]{
		\begin{minipage}[b]{0.4\textwidth}
			\includegraphics[width=1\textwidth]{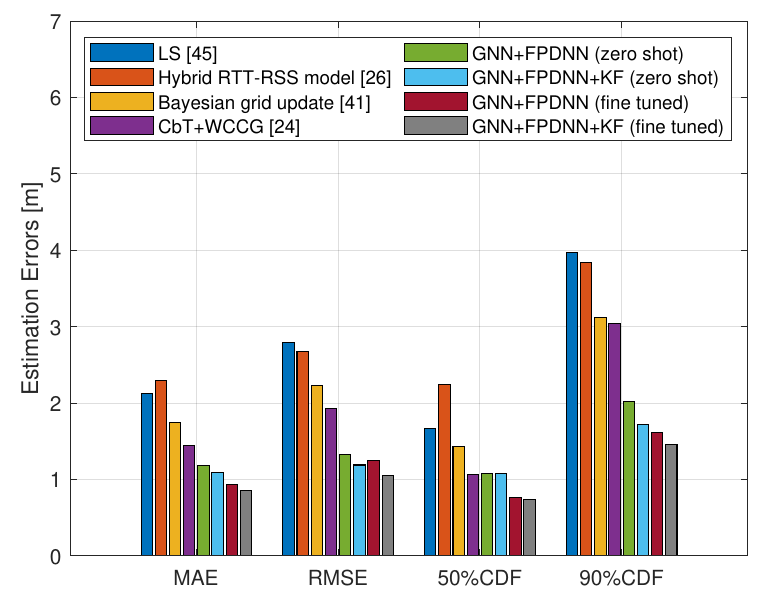} 
		\end{minipage}
			\quad

  \label{fig:shoppingmall_different_location_bar}
	}  
    	\subfigure[Estimation errors in the laboratory.]{
    		\begin{minipage}[b]{0.4\textwidth}
   		 	\includegraphics[width=1\textwidth]{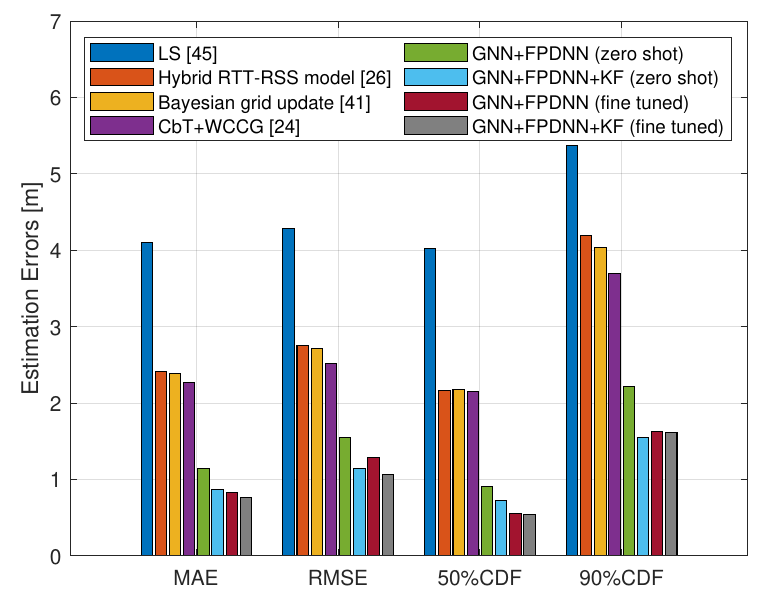}
    		\end{minipage}
		\label{fig:Lab_different_location_bar}
    	}
     \vspace{-0.1cm}
	\caption{Estimation errors of the proposed system and other baselines in different scenarios.}
	\label{fig:other_scenario experiment}
\end{figure*}

% \begin{figure}[htp]
% 	%\vspace{-0.1cm}
% 	\centering
% 	\begin{minipage}[t]{0.5\textwidth}
% 		\includegraphics[width=1\textwidth]{draft/different_propo_scenrio_fig_pdf.pdf}
% 	\end{minipage}
% 	%\vspace{-0.1cm}
% 	\caption{The location of APs in different scenarios. APs marked by asterisks for offline data collection and by squares for online data collection.}
% 	\label{fig:different_pro_scenario}
% 	\vspace{-0.3cm}
% \end{figure}

\subsection{Experimental Results in Different Scenarios} \label{different_location}
This part illustrates how to use our framework in unseen scenarios with no real-world data sample. 

The synthetic data generator and our localization algorithm are pre-trained with real-world data samples in the office environment. In the other two scenarios, shopping mall and laboratory, we do not use real-world data samples to fine-tune the generator and localization algorithm. To improve localization accuracy, the synthetic data generator exploits the floor-plan images to generate synthetic data in the new scenarios. The newly generated data samples are used to fine-tune the localization algorithm. As there is no real-world data sample, this approach is also referred to as zero-shot learning.

As a baseline, we can also fine-tune the localization algorithm with real-world data samples in the new scenarios. Like Fig.~\ref{fig:propo_scenario_online}, we fine-tune the GNN and FPDNN with one part of the samples and test them with the other part of the samples.

% , the GNN and FPDNN trained with synthetic data samples in the office are further tested in the .  in the new scenarios to fine-tune the synthetic data generator, GNN, or FPDNN.  and only exploit the floor-plan images in these scenarios to generate synthetic data samples.

% in different locations. Specifically, 
% by simply updating the floor plan of the new environment, we use the data generator pre-trained in Section~\ref{samearea} to generate the synthetic data. After the GNN and FPDNN are pre-trained by the synthetic data offline, we evaluate the performance of the system in the shopping mall. 

% During the fine-tuning stage, we use the first $50\%$ data samples collected in the shopping mall for fine-tuning the system and the last $50\%$ for testing. 

The MAE, RMSE, median, and $90$-th percentile CDF obtained in the shopping mall are provided in Fig.~\ref{fig:shoppingmall_different_location_bar}
Specifically, the RMSE achieved by zero-shot learning with (or without) the Kalman filter is $1.33$~m (or $1.19$~m). After fine-tuning, the RMSE is reduced to $1.25$~m (or $1.05$~m). The performance gap is around $6.0$\% $\sim 11.8$\%. We can observe similar performance gaps when using the other performance metrics. Nevertheless, the zero-shot learning approaches can reduce localization errors by around $31.1$\% $\sim 55.4$\% compared with the hybrid RTT-RSS, Bayesian grid update, and CbT+WCCG models.

In Fig.~\ref{fig:Lab_different_location_bar}, we estimate the localization errors in the laboratory. The performance gaps between zero-shot learning and fine-tuned approaches are around $7.0$\% $\sim 16.8$\%. Zero-shot learning approaches can reduce the localization errors by around $38.5$\% $\sim 58.7$\% compared with the hybrid RTT-RSS, Bayesian grid update, and CbT+WCCG models. We note that in Fig.~\ref{fig:Lab_different_location_bar}, the $90$-th percentile CDFs achieved by the zero-shot learning (GNN+FPDNN+KF) and the fined-tuned approach are $1.55$~m and $1.61$~m, respectively. This is because we use the MSE as the loss function to fine-tune the GNN and FPDNN, and this may result in higher peak errors.

\subsection{Multi-User Localization}\label{multi-user-condition}

\begin{figure}[htp]
	%\vspace{-0.1cm}
	\centering
	\begin{minipage}[t]{0.45\textwidth}
		\includegraphics[width=1\textwidth]{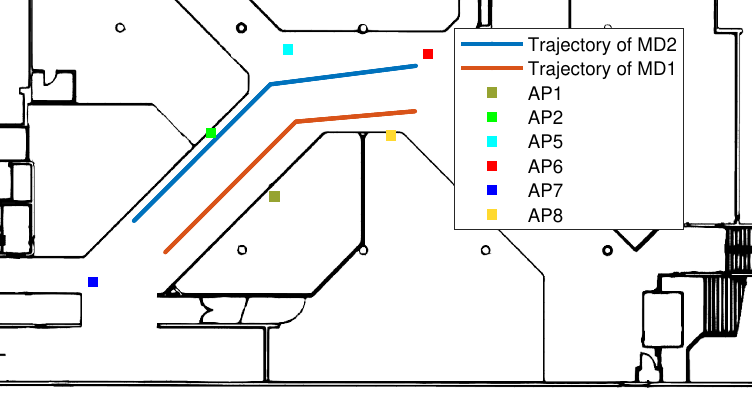}
	\end{minipage}
	\vspace{-0.1cm}
	\caption{Locations of two MDs and locations of six APs, where the MDs move along the trajectories.}
	\label{fig:shoppingmall_traj}
	
\end{figure}

This subsection evaluates our algorithm in multi-user scenarios, where $M=2$ MDs are considered. In our testing, MD~$1$ is connected to all APs, and MD~$2$ is connected to two of the APs. We only provide the performance obtained by the zero-shot learning approaches in the shopping mall. The gaps between the zero-shot learning and other baselines are similar to the single-user scenarios. Similar to Section~\ref{different_location}, we train the synthetic data generator in the office and use it to generate synthetic data samples in the shopping mall. The localization algorithm is only fine-tuned with the synthetic data samples, and real-world data samples in the shopping mall are only used for testing. For the laboratory scenario, there is no stable connection between the two MDs due to the blockages of walls and stairs. Thus, it is reduced to single-user localization. The testing samples of the two MDs are shown in Fig.~\ref{fig:shoppingmall_traj}. 

The results are provided in Table~\ref{tab:two_MD}. With the existing baselines, we first estimate the location of MD~$1$ and then use it to estimate the location of MD~$2$. Thus, the estimation error of MD~$1$ is accumulated in the estimation error of MD~$2$. This issue may have more significant impacts on localization accuracy when there are more than two MDs. Specifically, with the trilateration method, the RMSE gap between MD~$1$ and MD~$2$ is $1.18$~m. By using our GNN, we can obtain the estimated locations of the two MDs directly. With the GNN and Kalman filter, the RMSE gap between MD~$1$ and MD~$2$ is $0.53$~m. This observation implies that our approach can alleviate the accumulation of estimate errors. In addition, compared with the localization accuracy in the single-user scenario in Fig.~\ref{fig:shoppingmall_different_location_bar}, MD~$1$ achieves lower localization errors in the multi-user scenario. For example, the RMSE of MD~$1$ is $18.8$\% lower than the single-user scenario. This result indicates that the RTT and RSS information of MD~$2$ helps to improve the localization of MD~$1$ with our approach.

\begin{table}[]
 \vspace{-0.3cm}
\centering
\caption{Estimation Errors in The Shopping Mall with Two MDs}
\label{tab:two_MD}
\resizebox{85mm}{!}{%
\begin{tabular}{|c|c|c|c|c|c|}
\hline
                     & \textbf{Methods}  & \textbf{MAE} & \textbf{RMSE} & \textbf{50\% CDF} & \textbf{90\% CDF} \\ \hline
\multirow{5}{*}{MD1} & Trilateration     & 2.19         & 2.64          & 2.02              & 3.57              \\ \cline{2-6} 
                     & With GNN          & 1.22         & 1.57          & 1.05              & 1.75              \\ \cline{2-6} 
                     & With GNN+KF       & 1.03         & 1.23          & 0.92              & 1.71              \\ \cline{2-6} 
                     & With GNN+FPDNN    & 0.9          & 1.08          & 0.78              & 1.65              \\ \cline{2-6} 
                     & With GNN+FPDNN+KF & 0.79         & 0.9           & 0.71              & 1.37              \\ \hline
\multirow{5}{*}{MD2} & Trilateration     & 2.86         & 3.82          & 2.40              & 5.05              \\ \cline{2-6} 
                     & With GNN          & 1.86         & 2.47          & 1.39              & 3.54              \\ \cline{2-6} 
                     & With GNN+KF       & 1.42         & 1.76          & 1.19              & 2.52              \\ \cline{2-6} 
                     & With GNN+FPDNN    & 1.77         & 2.39          & 1.28              & 3.52              \\ \cline{2-6} 
                     & With GNN+FPDNN+KF & 1.47         & 1.7           & 1.37              & 2.47              \\ \hline
\end{tabular}%
}
 \vspace{-0.3cm}
\end{table}
\subsection{Complexity Analysis}
Since the GNN and FPDNN are trained offline, the training complexity has no impact on the localization performance. We analyze the complexity of online inference. The number of on-device parameters and computational complexity, which are evaluated by the number of floating point operations (FLOPs)
for processing one data sample of the proposed GNN and FPDNN, are provided in Table~\ref{tab:comm}. It can be seen that the numbers of FLOPs and on-device parameters of the proposed GNN model are $2,704$ and $1,381$, respectively. For the proposed FPDNN, the numbers of FLOPs and on-device parameters are $23,522,913$ and $863,097$, respectively.

We evaluate the average computation time for inference using the GNN and FPDNN with the NVIDIA GeForce RTX 3090. The processing time of the GNN and FPDNN is $2$~ms and $6$~ms, respectively. The existing baselines require more computation time on the GPU than the CPU. This is because GPUs are optimized for specialized computations, such as deep learning. Thus, we evaluate their processing time on the Intel(R) Core(TM) i9-12900KF CPU. Specifically, the processing time of the hybrid RTT-RSS model, the Bayesian grid update method, and the CbT+WCCG method are $2.94$~ms, $0.43$~ms, and $26$~ms, respectively. Compared with the minimum sampling interval between MDs and APs (200 ms), the processing time of all the localization algorithms is acceptable.

\begin{table}[]
 
\centering

\caption{The on-device parameters and computational complexity for proposed GNN and FPDNN}
\small
\label{tab:comm}
\resizebox{85mm}{!}{
\begin{tabular}{|c|c|c|}
\hline
\textbf{} & \textbf{On-device computation (FLOPs)} & \textbf{On-device parameters} \\ \hline
GNN & 2,704 & 1,381 \\ \hline
FPDNN & 23,522,913 & 863,097 \\ \hline
\end{tabular}
}

\end{table}

\section{Conclusions}
This paper proposed a multi-user indoor localization framework using floor-plan images, RTT distances, and RSS for localization. We first estimated the coarse locations of MDs by the GNN. Then, the positioning accuracy was further improved by the FPDNN. To improve the sample efficiency in real-world environments, we pre-trained the GNN and FPDNN with synthetic data samples generated in a virtual environment and tested them in different real-world environments, where no real-world data sample is available for training or fine-tuning the GNN and FPDNN. To collect data samples and validate our framework, we built a prototype, where the estimated locations of MDs are updated every $200$~ms. Our experimental results showed that the proposed indoor localization system significantly outperforms the existing baseline methods in terms of localization errors. Specifically, the zero-shot learning approaches can reduce localization errors by around $30$\% $\sim 55$\% compared with three baselines from the existing literature.
% with an RMSE of $0.29$ $m$. We also proposed a GNN model for multi-user localization, while some MDs can not obtain enough RTT distance from APs to perform accurate localization. We proved that information between each MD can
%further improve the localization accuracy of all the MDs with the GNN model. Besides, the GNN model can further enhance the localization accuracy of the MD estimated by the FPDNN model.

\ifCLASSOPTIONcaptionsoff
  \newpage
\fi

\bibliographystyle{IEEEtran}
\bibliography{main}

\begin{thebibliography}{10}
\providecommand{\url}[1]{#1}
\csname url@samestyle\endcsname
\providecommand{\newblock}{\relax}
\providecommand{\bibinfo}[2]{#2}
\providecommand{\BIBentrySTDinterwordspacing}{\spaceskip=0pt\relax}
\providecommand{\BIBentryALTinterwordstretchfactor}{4}
\providecommand{\BIBentryALTinterwordspacing}{\spaceskip=\fontdimen2\font plus
\BIBentryALTinterwordstretchfactor\fontdimen3\font minus
  \fontdimen4\font\relax}
\providecommand{\BIBforeignlanguage}[2]{{%
\expandafter\ifx\csname l@#1\endcsname\relax
\typeout{** WARNING: IEEEtran.bst: No hyphenation pattern has been}%
\typeout{** loaded for the language `#1'. Using the pattern for}%
\typeout{** the default language instead.}%
\else
\language=\csname l@#1\endcsname
\fi
#2}}
\providecommand{\BIBdecl}{\relax}
\BIBdecl

\bibitem{enge1994global}
P.~K. Enge, ``{The Global Positioning System: Signals, measurements, and
  performance},'' \emph{Int. J. Wireless Inf. Networks}, vol.~1, pp. 83--105,
  1994.

\bibitem{ozsoy2013indoor}
K.~Ozsoy, A.~Bozkurt, and I.~Tekin, ``Indoor positioning based on global
  positioning system signals,'' \emph{Microwave Opt. Technol. Lett.}, vol.~55,
  no.~5, pp. 1091--1097, 2013.

\bibitem{zhu2020indoor}
X.~Zhu, W.~Qu, T.~Qiu, L.~Zhao, M.~Atiquzzaman, and D.~O. Wu, ``{Indoor
  Intelligent Fingerprint-Based Localization: Principles, Approaches and
  Challenges},'' \emph{IEEE Commun. Surv. Tutorials}, vol.~22, no.~4, pp.
  2634--2657, 2020.

\bibitem{zafari2019survey}
F.~Zafari, A.~Gkelias, and K.~K. Leung, ``{A Survey of Indoor Localization
  Systems and Technologies},'' \emph{IEEE Commun. Surv. Tutorials}, vol.~21,
  no.~3, pp. 2568--2599, 2019.

\bibitem{he2015wi}
S.~He and S.-H.~G. Chan, ``{{Wi-Fi} Fingerprint-Based Indoor Positioning:
  Recent Advances and Comparisons},'' \emph{IEEE Commun. Surv. Tutorials},
  vol.~18, no.~1, pp. 466--490, 2016.

\bibitem{jianyong2014rssi}
Z.~Jianyong, L.~Haiyong, C.~Zili, and L.~Zhaohui, ``{{RSSI} based {B}luetooth
  Low Energy Indoor Positioning},'' in \emph{Proc. IEEE Int. Conf. Indoor
  Positioning Indoor Navig. (IPIN)}, 2014, pp. 526--533.

\bibitem{saab2010standalone}
S.~S. Saab and Z.~S. Nakad, ``{A Standalone {RFID} Indoor Positioning System
  Using Passive Tags},'' \emph{IEEE Trans. Ind. Electron.}, vol.~58, no.~5, pp.
  1961--1970, 2011.

\bibitem{zhang2016litell}
C.~Zhang and X.~Zhang, ``{LiTell: Robust Indoor Localization Using Unmodified
  Light Fixtures},'' in \emph{Proc. ACM Int. Conf. Mobile Comput. Networking
  (MOBICOM)}, 2016, pp. 230--242.

\bibitem{werner2011indoor}
M.~Werner, M.~Kessel, and C.~Marouane, ``Indoor positioning using smartphone
  camera,'' in \emph{Proc. IEEE Int. Conf. Indoor Positioning Indoor Navig.
  (IPIN)}, 2011, pp. 1--6.

\bibitem{8292759}
Z.~Ma, S.~Poslad, S.~Hu, and X.~Zhang, ``{A fast path matching algorithm for
  indoor positioning systems using magnetic field measurements},'' in
  \emph{Proc. IEEE 28th Annu. Int. Symp. Pers. Indoor, Mobile Radio Commun.
  (PIMRC)}, 2017, pp. 1--5.

\bibitem{8115912}
C.~Ma, C.~Wan, Y.~W. Chau, S.~M. Kang, and D.~R. Selviah, ``{Subway Station
  Real-time Indoor Positioning System for Cell Phones},'' in \emph{Proc. Int.
  Conf. Indoor Positioning Indoor Navig. (IPIN)}, 2017, pp. 1--7.

\bibitem{8647251}
Z.~Ma, J.~Bigham, S.~Poslad, B.~Wu, X.~Zhang, and E.~Bodanese, ``{Device-Free,
  Activity During Daily Life, Recognition Using a Low-Cost Lidar},'' in
  \emph{Proc. IEEE Global Commun. Conf. (GLOBECOM)}, 2018, pp. 1--6.

\bibitem{li2023cooperative}
X.~Li \emph{et~al.}, ``{A Cooperative Relative Localization System for
  Distributed Multi-Agent Networks},'' \emph{IEEE Trans. Veh. Technol.}, pp.
  1--16, 2023.

\bibitem{10.1145/3330180.3330193}
S.~u. Rahman, S.~Ullah, and S.~Ullah, ``{A Mobile Camera Based Navigation
  System for Visually Impaired People},'' in \emph{Proc. ACM 7th Int. Conf. on
  Commu. Broadband Networking (ICCBN)}, 2019, pp. 63--–66.

\bibitem{glanzer2009semi}
G.~Glanzer, T.~Bernoulli, T.~Wiessflecker, and U.~Walder, ``{Semi-autonomous
  Indoor Positioning Using MEMS-based Inertial Measurement Units and Building
  Information},'' in \emph{Proc. 6th Workshop on Positioning, Navig. and
  Commun. (WPNC)}, 2009, pp. 135--139.

\bibitem{9378557}
Y.~Yu, R.~Chen, L.~Chen, W.~Li, Y.~Wu, and H.~Zhou, ``{Autonomous 3D Indoor
  Localization Based on Crowdsourced Wi-Fi Fingerprinting and MEMS Sensors},''
  \emph{IEEE Sens. J.}, vol.~22, no.~6, pp. 5248--5259, 2022.

\bibitem{S24}
\BIBentryALTinterwordspacing
{Samsung S24 ultra}. Accessed: Mar. 2024. [Online]. Available:
  \url{{https://www.samsung.com/au/smartphones/galaxy-s24-ultra/}}
\BIBentrySTDinterwordspacing

\bibitem{Pixel}
\BIBentryALTinterwordspacing
{Google Pixel 8 Pro}. Accessed: Mar. 2024. [Online]. Available:
  \url{{https://store.google.com/au/product/pixel\_8\_pro}}
\BIBentrySTDinterwordspacing

\bibitem{singh2021machine}
N.~Singh, S.~Choe, and R.~Punmiya, ``{Machine Learning Based Indoor
  Localization Using Wi-Fi RSSI Fingerprints: An Overview},'' \emph{IEEE
  Access}, vol.~9, pp. 127\,150--127\,174, 2021.

\bibitem{basri2016survey}
C.~Basri and A.~El~Khadimi, ``{Survey on Indoor localization System and Recent
  Advances of WIFI Fingerprinting Technique},'' in \emph{Proc. IEEE Int. Conf.
  Multimedia Comput. Syst. (ICMCS)}, 2016, pp. 253--259.

\bibitem{zheng2023exploiting}
C.~Zhou, J.~Liu, M.~Sheng, Y.~Zheng, and J.~Li, ``{Exploiting Fingerprint
  Correlation for Fingerprint-Based Indoor Localization: A Deep Learning Based
  Approach},'' \emph{IEEE Trans. Veh. Technol.}, vol.~70, no.~6, pp.
  5762--5774, 2021.

\bibitem{ibrahim2018verification}
M.~Ibrahim \emph{et~al.}, ``{Verification: Accuracy Evaluation of WiFi Fine
  Time Measurements on an Open Platform},'' in \emph{Proc. ACM Int. Conf. on
  Mobile Comput. and Networking (MOBICOM)}, 2018, pp. 417--427.

\bibitem{gentner2020wifi}
C.~Gentner, M.~Ulmschneider, I.~Kuehner, and A.~Dammann, ``{{WiFi-RTT} Indoor
  Positioning},'' in \emph{Proc. IEEE Symp. Position Locat. Navig. (PLANS)},
  2020, pp. 1029--1035.

\bibitem{9151400}
C.~Ma, B.~Wu, S.~Poslad, and D.~R. Selviah, ``{Wi-Fi RTT Ranging Performance
  Characterization and Positioning System Design},'' \emph{IEEE Trans. Mob.
  Comput.}, vol.~21, no.~2, pp. 740--756, 2022.

\bibitem{martin2020ranging}
I.~Martin-Escalona and E.~Zola, ``{Ranging Estimation Error in WiFi Devices
  Running IEEE 802.11mc},'' in \emph{Proc. IEEE Conf. Global Commun.
  (GLOBECOM)}, 2020, pp. 1--7.

\bibitem{8924707}
G.~Guo, R.~Chen, F.~Ye, X.~Peng, Z.~Liu, and Y.~Pan, ``{Indoor Smartphone
  Localization: A Hybrid WiFi RTT-RSS Ranging Approach},'' \emph{IEEE Access},
  vol.~7, pp. 176\,767--176\,781, 2019.

\bibitem{dvorecki2019machine}
N.~Dvorecki, O.~Bar-Shalom, L.~Banin, and Y.~Amizur, ``{A Machine Learning
  Approach for Wi-Fi RTT Ranging},'' in \emph{Proc. Int. Tech. Meet. Inst.
  Navig.}, 2019, pp. 435--444.

\bibitem{wang2020nlos}
Y.~Wang, K.~Gu, Y.~Wu, W.~Dai, and Y.~Shen, ``{NLOS Effect Mitigation via
  Spatial Geometry Exploitation in Cooperative Localization},'' \emph{IEEE
  Trans. Wireless Commun.}, vol.~19, no.~9, pp. 6037--6049, 2020.

\bibitem{huang2020hpips}
L.~Huang \emph{et~al.}, ``{HPIPS: A High-Precision Indoor Pedestrian
  Positioning System Fusing WiFi-RTT, MEMS, and Map Information},''
  \emph{Sensors}, vol.~20, no.~23, p. 6795, 2020.

\bibitem{9293264}
H.~Cao, Y.~Wang, and J.~Bi, ``{Smartphones: 3D Indoor Localization Using Wi-Fi
  RTT},'' \emph{IEEE Commun. Lett.}, vol.~25, no.~4, pp. 1201--1205, 2021.

\bibitem{si2020wi}
M.~Si, Y.~Wang, S.~Xu, M.~Sun, and H.~Cao, ``{A Wi-Fi FTM-Based Indoor
  Positioning Method with LOS/NLOS Identification},'' \emph{Appl. Sci.},
  vol.~10, no.~3, p. 956, 2020.

\bibitem{cao2020indoor}
H.~Cao, Y.~Wang, J.~Bi, S.~Xu, M.~Si, and H.~Qi, ``{Indoor Positioning Method
  Using WiFi RTT Based on LOS Identification and Range Calibration},''
  \emph{ISPRS Int. J. of Geo-Inf.}, vol.~9, no.~11, p. 627, 2020.

\bibitem{han2019smartphone}
K.~Han, S.~M. Yu, and S.-L. Kim, ``{Smartphone-based Indoor Localization Using
  Wi-Fi Fine Timing Measurement},'' in \emph{Proc. IEEE Int. Conf. Indoor
  Positioning Indoor Navig. (IPIN)}, 2019, pp. 1--5.

\bibitem{bregar2018improving}
K.~Bregar and M.~Mohor{\v{c}}i{\v{c}}, ``{Improving Indoor Localization Using
  Convolutional Neural Networks on Computationally Restricted Devices},''
  \emph{IEEE Access}, vol.~6, pp. 17\,429--17\,441, 2018.

\bibitem{dong2021real}
Y.~Dong, T.~Arslan, and Y.~Yang, ``{Real-Time NLOS/LOS Identification for
  Smartphone-Based Indoor Positioning Systems Using WiFi RTT and RSS},''
  \emph{IEEE Sens. J.}, vol.~22, no.~6, pp. 5199--5209, 2021.

\bibitem{choi2022enhanced}
J.~Choi, ``{Enhanced Wi-Fi RTT Ranging: A Sensor-Aided Learning Approach},''
  \emph{IEEE Trans. Veh. Technol.}, vol.~71, no.~4, pp. 4428--4437, 2022.

\bibitem{li2023variational}
Y.~Li, S.~Mazuelas, and Y.~Shen, ``{A Variational Learning Approach for
  Concurrent Distance Estimation and Environmental Identification},''
  \emph{IEEE Trans. Wireless Commun.}, vol.~22, no.~9, pp. 6252--6266, 2023.

\bibitem{seong2021high}
J.-H. Seong, S.-H. Lee, W.-Y. Kim, and D.-H. Seo, ``{High-Precision RTT-Based
  Indoor Positioning System Using RCDN and RPN},'' \emph{Sensors}, vol.~21,
  no.~11, p. 3701, 2021.

\bibitem{guo2023factor}
G.~Guo, R.~Chen, X.~Niu, K.~Yan, S.~Xu, and L.~Chen, ``{Factor Graph Framework
  for Smartphone Indoor Localization: Integrating Data-Driven PDR and Wi-Fi
  RTT/RSS Ranging},'' \emph{IEEE Sens. J.}, vol.~23, no.~11, pp.
  12\,346--12\,354, 2023.

\bibitem{zehua2023indoor}
L.~Zehua, S.~Junna, and Y.~Ziyang, ``{Indoor Integrated Navigation on
  PDR/Wi-Fi/barometer via Factor Graph with Local Attention},'' \emph{IEEE
  Sens. J.}, pp. 1--1, 2023.

\bibitem{horn2022indoor}
B.~Horn, ``{Indoor Localization Using Uncooperative Wi-Fi Access Points},''
  \emph{Sensors}, vol.~22, no.~8, p. 3091, 2022.

\bibitem{dosovitskiy2020image}
\BIBentryALTinterwordspacing
A.~Dosovitskiy \emph{et~al.}, ``{An Image is Worth 16x16 Words: Transformers
  for Image Recognition at Scale},'' 2021, \textit{arXiv:2010.11929}. [Online].
  Available: \url{https://arxiv.org/abs/2010.11929}
\BIBentrySTDinterwordspacing

\bibitem{zhou2021deepvit}
\BIBentryALTinterwordspacing
D.~Zhou \emph{et~al.}, ``{DeepViT: Towards Deeper Vision Transformer},'' 2021,
  \textit{arXiv:2103.11886}. [Online]. Available:
  \url{https://arxiv.org/abs/2103.11886}
\BIBentrySTDinterwordspacing

\bibitem{bottou2010large}
L.~Bottou, ``{Large-Scale Machine Learning with Stochastic Gradient Descent},''
  in \emph{Proc. Int. Conf. Comput. Stat. (COMPSTAT)}, 2010, pp. 177--186.

\bibitem{4212819}
I.~Guvenc, C.-C. Chong, and F.~Watanabe, ``{Analysis of a Linear Least-Squares
  Localization Technique in LOS and NLOS Environments},'' in \emph{Proc. IEEE
  Veh. Technol. Conf. (VTC)}, 2007, pp. 1886--1890.

\bibitem{9459462}
M.~Lecci, P.~Testolina, M.~Polese, M.~Giordani, and M.~Zorzi, ``{Accuracy
  Versus Complexity for mmWave Ray-Tracing: A Full Stack Perspective},''
  \emph{IEEE Trans. Wirel. Commun.}, vol.~20, no.~12, pp. 7826--7841, 2021.

\bibitem{kingma2014adam}
\BIBentryALTinterwordspacing
D.~P. Kingma and J.~Ba, ``{Adam: A Method for Stochastic Optimization},'' 2014,
  \textit{arXiv:1412.6980}. [Online]. Available:
  \url{https://arxiv.org/abs/1412.6980}
\BIBentrySTDinterwordspacing

\end{thebibliography}

\end{document}